\documentclass[aps,prb,twocolumn,showpacs,superscriptaddress,longbibliography]{revtex4-1}  
\usepackage{graphicx}  
\usepackage{dcolumn}   
\usepackage{bm}        
\usepackage{amsmath}
\usepackage{amsthm}
\usepackage{amssymb}
\usepackage{bbold} 
\usepackage{mathrsfs}
\usepackage{array}
\usepackage[dvipsnames,usenames]{color}
\usepackage{soul} 
\usepackage{ulem} 
\usepackage{physics}
\usepackage{ulem}

\usepackage{xcolor}
\usepackage[dvipsnames,usenames]{color}

\usepackage{braket}
\usepackage{float}
\usepackage{mhchem}
\usepackage[makeroom]{cancel}
\usepackage{upgreek}
\usepackage{commath}
\usepackage{siunitx}

\usepackage[titletoc]{appendix}

\usepackage{hyperref}
\hypersetup{colorlinks=true, linkcolor=magenta, citecolor=magenta, urlcolor=magenta}

\begin{document}
\title {Multifractal critical phase driven by coupling quasiperiodic systems to electromagnetic cavities}

\author{Thales F. Macedo}
\affiliation{Instituto de F\'isica, Universidade Federal do Rio de Janeiro, Cx.P. 68.528, 21941-972 Rio de Janeiro RJ, Brazil}
\author{Juli\'an Fa\'undez}
\affiliation{Instituto de F\'isica, Universidade Federal do Rio de Janeiro, Cx.P. 68.528, 21941-972 Rio de Janeiro RJ, Brazil}
\affiliation{Departamento de F\'isica y Astronomía, Universidad Andres Bello, Santiago 837-0136, Chile}
\author{Raimundo R.~dos Santos}
\affiliation{Instituto de F\'isica, Universidade Federal do Rio de Janeiro, Cx.P. 68.528, 21941-972 Rio de Janeiro RJ, Brazil}
\author{Natanael C.~Costa}
\affiliation{Instituto de F\'isica, Universidade Federal do Rio de Janeiro, Cx.P. 68.528, 21941-972 Rio de Janeiro RJ, Brazil}
\author{Felipe A. Pinheiro}
\affiliation{Instituto de F\'isica, Universidade Federal do Rio de Janeiro, Cx.P. 68.528, 21941-972 Rio de Janeiro RJ, Brazil}


\begin{abstract} 
We theoretically investigate criticality and multifractal states in a one-dimensional Aubry-Andre-Harper model coupled to electromagnetic cavities. 
We focus on two specific cases where the phonon frequencies are $\omega_{0}=1$ and $\omega_{0}=2$, respectively. 
Phase transitions are analyzed using both the average and minimum inverse participation ratio to identify metallic, fractal, and insulating states.
We provide numerical evidence to show that the presence of the optical cavity induces a critical, intermediate phase in between the extended and localized phases, hence drastically modifying the traditional transport phase diagram of the Aubry-Andre-Harper model, in which critical states can only exist at the well-defined metal-insulator critical point. We also investigate the probability distribution of the inverse participation ratio and conduct a multifractal analysis to characterize the nature of the critical phase, in which we show that extended, localized, and fractal eigenstates coexist. Altogether our findings reveal the pivotal role that the coupling to electromagnetic cavities plays in tailoring critical transport phenomena at the microscopic level of the eigenstates.
\end{abstract}

\maketitle

\section{Introduction}

The study of systems with strong light-matter interactions has gained significant interest over the past years, particularly because it offers new avenues for understanding fundamental properties of matter\,\cite{Ritsch2013,Schlawin2022}. In the strong coupling regime, the behavior of dressed quasiparticles and their collective modes are affected beyond the rotating-wave approximation or mean-field theory, which may lead to new states of matter or unexpected phenomena. 
Indeed, with the advent of optical cavities with high-quality factors, a wide range of experiments involving strong light-matter coupling have become possible.
For instance, cavity quantum electrodynamics (QED) experiments have demonstrated that the coupling with the electromagnetic field can lead to supersolid\,\cite{Leonard2017a,Leonard2017b} and superradiant Mott insulating phases\,\cite{Landig2016,Klinder2015,Vaidya2018} in Bose-Einstein condensates. 
Additionally, Ref.\,\cite{Budden2021} used mid-infrared laser pulses to generate light-induced superconducting states with nanoseconds lifetime in K$_{3}$C$_{60}$ compounds, leading to theoretical developments that could enable cavity-enhanced superconductivity\, \cite{Sentef2018,Schlawin2019,Curtis2019}.
Therefore, understanding how the interaction with photons drives to such phases is crucial for learning about their nature and, more importantly, how to manipulate them.

Within this context, a great deal of interest is especially focused on transport properties. For instance, quantum state transfer or state transfer protocols have been investigated on coupled cavity arrays\,\cite{Hartmann2006,Tomadin2010,Baum2022,Saxena2023,Patton2024}, an issue of much interest in quantum computing. In addition, semiconductor quantum detectors have shown an enhancement of their photoconductivity due to strong light-matter coupling (and its eventual collective effects)\,\cite{Pisani2023}, while the topological protection of the integer quantum Hall effect can be disrupted by long-range electron hopping induced by cavity QED fluctuations\,\cite{Appugliese2022}. 
Of particular interest to the present work are recent experiments on disordered organic semiconductors, where carrier states are hybridized with the electromagnetic field\,\cite{Orgiu2015}. These experiments have demonstrated an enhancement of the conductivity by an order of magnitude, suggesting that the coupling with photons can mitigate the harmful effects of randomness on electronic transport.

In view of these stimulating results, a great theoretical effort has been made over the past years, to understand the leading effects of non-perturbative cavity light-matter coupling. For instance, Ref.\,\cite{Hagenmuller2017} showed that the coupling with the cavity enhances the conductivity. 
Further, a Dicke superradiant quantum phase transition may emerge in a cavity coupled with a 2D electron gas with  Rashba spin-orbit interaction \cite{Nataf2019}.
Other important theoretical developments were also made for topological phases\,\cite{Wang2019,Guerci2020,Jangjan2020,Dmytruk2022,Allard23,Ezawa2024,Liu2024,Nguyen2024}, Majorana fermions\,\cite{Bacciconi24,Leon2024}, entanglement properties\,\cite{Passetti23},
magnetic or superconducting properties\,\cite{Ashida2020,Masuki24,Li2020}, Kondo effect\,\cite{Mochida2024}, and disordered systems\,\cite{Arwas2023,Moreno2022,Guo2024}. In particular, Ref.~\cite{Moreno2022} investigated the effects of cavity QED in an insulating 1D disordered chain, demonstrating that conductivity can be enhanced by several orders of magnitude due to the emission and absorption of virtual photons. This implies that the localization length is highly dependent on the coupling with the cavity~\cite{Moreno2022}. Since the 1D Anderson model does not exhibit a metal-insulator transition, these results only apply to exponentially localized, Anderson wavefunctions in disordered media. However, the impact of the coupling to electromagnetic cavities on metal-insulator transitions remains unknown.
 
To address this issue, we consider a simple Hamiltonian exhibiting extended-localized phase transition, the Aubry-Andre-Harper (AAH) model\,\cite{Harper55,Aubry80}. 
It describes a one-dimensional chain with a quasiperiodic on-site potential, mimicking a quasicrystal. That is, the potential is not random, as in the Anderson model, but aperiodic, as required for extended Bloch states\,\cite{Aulbach2004,Bu2022}. As discussed below, the AAH model is notable for being self-dual, thus having extended-localized phase transitions to \textit{all} states at the same critical point \cite{Dominguez2019}. Therefore, the main goal of this paper is to investigate how the cavity QED affects the critical properties of the AAH model, particularly in the strong coupling regime.

We analyze the cavity QED Aubry-Andre-Harper model through  exact diagonalization methods; localization properties are probed by the behavior of inverse participation ratio and multifractal analysis.
We show that the critical point that exists in the absence of the cavity changes into an entire critical phase where critical, extended, and localized eigenstates coexist, which broadens with increasing cavity coupling. 

The paper is organized as follows.  The AAH model, its coupling with the electromagnetic field of the cavity and the quantities of interest are presented in Section\,\ref{Sec:model}. In Section \ref{Sec:results} we present and discuss our main results, while our conclusions and further remarks are left to Section\,\ref{conclusion}.

\begin{figure}
    \centering
    \includegraphics[scale=0.70]{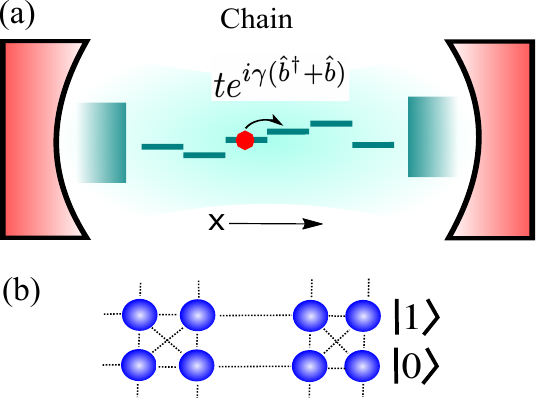}
    \caption{ (a) Schematic representation of a chain inside an optical cavity. 
    The horizontal lines lie on the chain sites, and denote photon levels, whose height indicates the number, $n$, of electromagnetic quanta.
    $\gamma$ is the coupling strength with the electromagnetic field within the cavity. (b) As the electron hops, it can absorb or emit a photon, effectively leading to a multi-chain perspective, with $\ket{n}$ denoting the photon state.}
    \label{Fig:esquema}
\end{figure}

\section{Model and methodology}\label{Sec:model}

\subsection{The model}\label{subsection:model}

The one-dimensional AAH model describes spinless fermions under a quasiperiodic potential\,\cite{Harper55,Aubry80}. 
Its Hamiltonian reads
\begin{align}
\nonumber
\mathcal{H}_{AA} = &-t\sum_{\mathbf{i} }(c^{\dagger}_{\mathbf{i}}c_{\mathbf{i} + \mathbf{\hat{x}}}^{\phantom{\dagger}}+ \text{H.c}) \\
& +  g_{AA}\sum_{\mathbf{i}}\cos{( 2\pi \beta |\mathbf{i}| + \phi)}c^{\dagger}_{\mathbf{i}}c_{\mathbf{i}}^{\phantom{\dagger}}, 
    \label{Eq:Hamiltonian1}
\end{align}
where the sums run over a one-dimensional chain.
Here, we use the standard second quantization formalism, in which $c_{\mathbf{i}}^\dagger$ ($c_{\mathbf{i}}$) are creation (annihilation) operators of fermions at a given site $\mathbf{i}$.
The first term on the right-hand side of Eq.\,\eqref{Eq:Hamiltonian1} describes the fermionic hopping, while the second term corresponds to the onsite potential, with strenght $g_{AA}$.
To describe incommensurate features, one may deal with open boundary conditions (OPC) and define $\beta=\frac{\sqrt{5}-1}{2}$, i.e.~as the inverse golden ratio, with $\phi$ being an arbitrary global phase.
For periodic boundary conditions (PBC), it is achieved by defining $\beta$ as the ratio of two adjacent Fibonacci numbers, $F_{m-1}/F_{m}$, and extrapolating the results to the thermodynamic limit.
Hereafter, the hopping integral, $t$, sets the energy scale and may be taken as unity; similarly, we also take the lattice constant, $a$, as unity.  

Interestingly, the Aubry-Andre-Harper model is self-dual, \textit{i.e.}~it is symmetric under a Fourier transform, leading to a sharp transition between extended and localized states: \textit{all} eigenstates are extended (localized) for $g_{AA} < 2$ ($g_{AA} > 2$).
The model exhibits critical properties only at $g_{AA} = 2$, at which the wave function is multifractal \cite{Dominguez2019}.
Many different extensions of the AAH model have been investigated over the past decades, with the emergence of edge states\,\cite{Biddle2011,Ganeshan2015,Liu2022}, as well as for higher dimensions\,\cite{Rossignolo2019}.
However, examining cavity effects on these extended models is beyond the scope of this work. 

As mentioned in the Introduction, we place our one-dimensional AAH electronic system within a cavity with a single-mode electromagnetic field.
Following Refs.\,\cite{Moreno2022} and \cite{Jiajun2020}, the coupling between the bosonic and fermionic degrees of freedom changes the hopping integral (as a Peierls substitution)\,\cite{Peierls1933,Marder2010}, $t \to t \exp\big[ i e a A_{x} / \hbar c\big]$, with $A_{x}$ being the vector potential of the electromagnetic field, and $e$ the fermion charge.
Here we assume that the wavelength is much larger than the lattice spacing, leading to a position-independent vector potential.
Given this, we define $A_{x} = A_{0} (\hat{b}+\hat{b}^{\dagger})$, with $\hat{b}^{\dagger}$ ($\hat{b}$) being creation (annihilation) operators of photons of frequency $\omega_{0}$.
This is illustrated in Fig.\,\ref{Fig:esquema}\,(a), where the renormalized (Peierls) electronic hopping connects states with different photon numbers through absorption or emission processes.
The cavity-coupled Hamiltonian then reads, 
\begin{align}
\nonumber
\mathcal{H} = &-t\sum_{\mathbf{i}}\bigg[ e^{i \gamma(\hat{b}+\hat{b}^{\dagger})} c^{\dagger}_{\mathbf{i}}c_{\mathbf{i} + \mathbf{\hat{x}}}^{\phantom{\dagger}}+ \text{H.c}\bigg] \\
& +  g_{AA}\sum_{\mathbf{i}}\cos{( 2\pi \beta |\mathbf{i}| + \phi)}c^{\dagger}_{\mathbf{i}}c_{\mathbf{i}}^{\phantom{\dagger}} +\hbar\omega_{0}\hat{b}^{\dagger}\hat{b}, 
    \label{Eq:Hamiltonian2}
\end{align}
where $\gamma = e a A_{0}/(\hbar c)$ describes the coupling with the electromagnetic field, and the last term is the free photons contribution to the energy.

We investigate the Hamiltonian of Eq.\,\eqref{Eq:Hamiltonian2} through exact diagonalization methods in the subspace of a single fermion coupled to $N_{ph}$ photons.
A convenient basis for our Hilbert space is spanned by $\ket{l,n}$, with $1 \leq l \leq L$ labeling the site position, and $0 \leq n \leq N_{ph}$ being the photon quantum number.
Using the Baker-Campbell-Hausdorff formula,
the matrix elements for the Peierls operator can be expressed as
\begin{widetext}
\begin{align}
\langle n| e^{i\gamma(b+b^{\dagger})} |m \rangle = & e^{-\gamma^2/2} \sum_{s=0}^{n} \sum_{j=0}^{m}
\frac{(i\gamma)^s}{s!} \frac{(i\gamma)^j}{j!} 
 \times \delta_{n-s,m-j} P(s,n) P(j,m)~,
\end{align}
\end{widetext}
obtained by expanding the operators as a Taylor series, and with $P(j,m) = \sqrt{m [m-1] [m-2] \dots [(m - (j-1)]}$ (see, e.g., Ref.\,\cite{Moreno2022}). As mentioned before, it renormalizes the hopping integrals within the different photon processes. 
Given this, the matrix representation of the Hamiltonian is 
 $$\mathcal{H} =
 \begin{pmatrix}
  \hat{H}_{0,0} & \hat{H}_{0,1} & \dots & \hat{H}_{0,N_{ph}} \\ 
  \hat{H}_{1,0} & \hat{H}_{1,1} & \dots & \hat{H}_{1,N_{ph}} \\
  \vdots & \vdots & \dots & \vdots \\
  \hat{H}_{N_{ph},0} & \hat{H}_{N_{ph},1} & \dots & \hat{H}_{N_{ph},N_{ph}}
\end{pmatrix}$$
with $\hat{H}_{n,m}$ denoting $L \times L$ matrices whose matrix elements are $ \big[\hat{H}_{n,m} \big]_{i,j} = \langle i, n | \mathcal{H} | j, m \rangle $.
The final $(N_{ph} +1) \times L$ matrix is fully diagonalized by standard linear algebra numerical routines, which provide the eigenvalues and eigenvectors.

This scenario effectively corresponds to a problem involving strongly coupled chains, as illustrated in Fig.\,\ref{Fig:esquema}\,(b). In this representation, $\ket{n}$ denotes the photon number state, and the links between different photon states correspond to photon-assisted hopping terms. In words, the original system without photons, described by $\hat{H}_{0,0}$, is coupled to a series of virtual replicas -- each with $n$ photons, renormalized hopping amplitudes, and energy spectra shifted by $n \hbar \omega_0$. These replicas are represented by the matrices $\hat{H}_{n,n}$. The interaction between these chains arises through photon absorption and emission processes, which are captured by the off-diagonal matrices $\hat{H}_{n,m}$ for $n \neq m$.
In this work, we use both OBC and PBC, while averaging over 10 different random values of $\phi$.
The relatively small number of disorder configurations is justified by the weak dependence of the inverse participation ratio on $\phi$ in the AAH model, as confirmed \textit{a posteriori}.
We also set a cutoff for the total number of photons at $N_{ph}=8$, as discussed below.

\subsection{The inverse participation ratio} 

Using the basis states $\ket{l,n}$ for our Hilbert space, we may write a generic (normalized) eigenstate of $\mathcal{H}$ as
\begin{equation}
\label{Eq:eigenstate}
\ket{\psi_{j}}=\sum_{l,n}\phi_{l,n}^{(j)}\ket{l,n},
\end{equation}
where $j$ labels the number of the eigenstate, and $\phi_{l,n}^{(j)}$ is the probability amplitude of finding the system in $\ket{l,n}$.

Given this, one key quantity of interest is the inverse participation ratio (IPR), which determines if a given wavefunction is spatially localized.
It is defined as
\begin{equation}
    \text{IPR}(\ket{\psi_{j}}) = \sum_{l,n}|\phi_{l,n}^{(j)}|^{2 q}~,
\end{equation}
with $q=2$.
If $\text{IPR}(\ket{\psi_j}) \to 1$ in the limit $L \to \infty$, the state $\ket{\psi_j}$ is localized in a single site. On the other hand, if
$\text{IPR}(\ket{\psi_j}) \to 0$ in the limit $L \to \infty$, the state is extended, meaning that $\ket{\psi_j}$ is delocalized and spreads over all $L$ sites in the system.

\begin{figure}[t]
    \centering
   \includegraphics[scale=0.50]{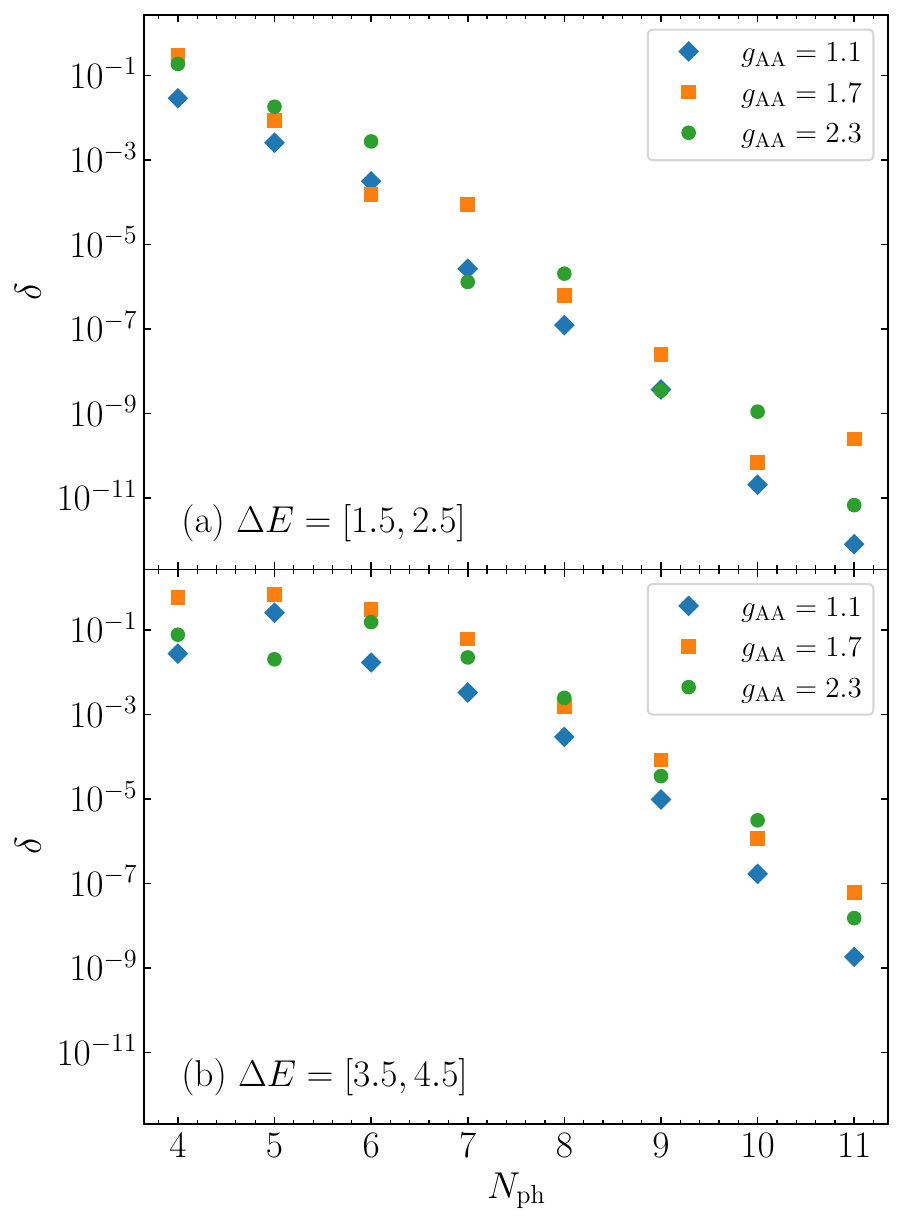}
     \caption{Log-linear plots of the relative $\overline{\text{IPR}}$ error [Eq.\,\eqref{Eq:delta}], averaged over the energy ranges (a) $\Delta E = [1.5, 2.5]$, and (b) $\Delta E = [3.5, 4.5]$, as functions of the number of photons, $N_{\text{ph}}$, for different values of the disorder strength, $g_{AA}$.}
    \label{Fig:delta}
\end{figure}

It is more convenient to examine the behavior of the IPR when averaged over a set of $N_{st}$ eigenstates within a specific energy window,
$\Delta E = [E_{M_i}, E_{M_f}]$, $M_i\leq j\leq M_f$,
\begin{equation}
    \overline{\text{IPR}} = \frac{1}{N_{st}}\sum_{j=M_i}^{M_f}\text{IPR}(\ket{\psi_{j}}).
    \label{eq:IPRbar}
\end{equation}
We note that the emission and absorption of photons couple different states across the photon-number subspaces, which would lead to inhomogeneities for the localization properties within a given energy window. Therefore, it is useful to define the typical IPR,
\begin{equation}
\text{IPR}_{\text{typ}} = e^{\langle \ln \text{IPR} \rangle} ~,
\end{equation}
with
\begin{equation}
\langle \ln \text{IPR} \rangle = \frac{1}{N_{st}}\sum_{j=M_i}^{M_f}\ln \big[ \text{IPR}(\ket{\psi_{j}}) \big]~.
\end{equation}
This quantity effectively distinguishes delocalized from localized states and provides a more accurate determination of the critical points.
Similarly, and within the same energy window, one may define the minimum IPR, that is
\begin{equation}
\text{IPR}_{min} = {\rm min}\big[\text{IPR}(\ket{\psi_{j}})\big], \forall ~ M_i \leq j \leq M_f~. 
\end{equation}
While  $\overline{\text{IPR}}$ and  $\text{IPR}_{\text{typ}}$ determine the occurrence of localized states for a given set of external parameters, $\text{IPR}_{min}$ establishes the threshold for the occurrence of at least one extended state.

\subsection{The cutoff in the number of photons $N_{ph}$} 

At this point, it is important to examine the cutoff of $N_{\text{ph}}$, which should be chosen such that it does not affect the accuracy of the physical quantities of the system. Since the different photon-number subspaces $\hat{H}_{n,n}$ are energetically shifted by $n\hbar \omega_{0}$, the impact of the cutoff depends on the specific energy window being investigated. Of particular interest are the low-energy states, due to their importance for transport properties. One might expect these states to be only weakly affected by contributions from subspaces with large number of photons, especially when $n\hbar\omega_0$ exceeds the bare electron bandwidth.

To determine the cutoff for the photon number $N_{\text{ph}}$, we analyze the behavior of $\overline{\text{IPR}}$ as a function of $N_{\text{ph}}$. This is done by comparing it to a reference case with a sufficiently large photon number, which serves as an approximation for the asymptotic behavior. To quantify the deviation from the reference, we compute the relative error in $\overline{\text{IPR}}$,
\begin{equation}
\delta (N_{\text{ph}}) = \frac{| \overline{\text{IPR}}(N_{\text{ph}}) - \overline{\text{IPR}}_{\text{ref}}|}{\overline{\text{IPR}}_{\text{ref}}}~,
\label{Eq:delta}
\end{equation}
with $\overline{\text{IPR}}_{\text{ref}}$ being the reference value taken from the case with $N_{\text{ph}} = 12$.
The behavior of $\delta (N_{\text{ph}})$ is shown in Fig.\,\ref{Fig:delta} for two different energy windows.
In both scenarios, and across different disorder strengths, the corrections to $\overline{\text{IPR}}$ become negligible once the photon number reaches $N_{\text{ph}} \geq 8$. Although higher-energy states might require more photons for an accurate representation, such states are generally less significant in the context of transport experiments. Therefore, in what follows, we fix the photon number to $N_{\text{ph}} = 8$.

\subsection{Fractal properties}
\label{MFA}

Another interesting feature of the IPR is its relation with the fractal dimension of the wavefunction, which indicates how distributed it is within the medium. 
For extended states, the probability of finding the particle at a given site is proportional to $1/L^{d}$, where $d$ is the dimensionality of the system. 
This leads to $\text{IPR} \approx L^{d(1-q)}$, which simplifies to $\text{IPR} \approx L^{-d}$ for $q=2$. 
However, at the critical point, the wavefunction coverage over the medium becomes fractal, rather than homogeneous; accordingly, the scaling properties of the IPR are described through the replacement $d \to D_f$, where $D_f$ is the fractal dimension.

In the presence of $N_{ph}$ photons, the extension of these ideas is straightforward.
For extended states, the wavefunction amplitudes are expected to scale as $|\phi_{l,n}^{j}|^{2} \approx A_n/L^{d}$, where $A_n$ is a normalization factor that depends on the photon subspace index $n$. Under this assumption, the IPR behaves as
$$
\text{IPR} \approx \left( \sum_{n=1}^{N_{ph}} A_n^q \right) L^{d(1-q)} = C_{N_{ph}} L^{d(1-q)}~,
$$
with $C_{N_{ph}} = \sum_{n=1}^{N_{ph}} A_n^q$ taking into account the contributions from different photon subspaces.
This leads to the scaling relation
\begin{align}
\label{Eq:fractal-dim}
\log(\text{IPR}) \approx -D_f \log(L) + \log(C_{N_{ph}}),
\end{align}
for $q = 2$.
Importantly, for large photon number cutoffs $N_{ph}$, the sum over $n$ in $C_{N_{ph}}$ is often dominated by a few photon subspaces, rendering $C_{N_{ph}}$ constant. As a result, $C_{N_{ph}}$ does not alter the scaling behavior of the IPR with system size, thus not affecting the extracted fractal dimension $D_f$.

Further features of the wavefunction are provided by performing a multifractal analysis \cite{Kohmoto1983,Liu2022}.
Assuming PBC, we choose system sizes from a 
Fibonacci sequence, $L=F_m$. As the probability of finding the particle on a given site $l$ is 
$P_l \propto F_m^{-\alpha_{l}^{(j)}}$, the set of exponents $\alpha_{l}^{(j)}$ are used to characterize 
the distribution of the state $\ket{\psi_{j}}$ over the sites, as follows.
We define $\alpha_{min}^{(j)} \equiv \text{min}\big[\alpha_{l}^{(j)}\big]$, with $1\leq l \leq F_m$, and take $m\to\infty$: if $\alpha_{min}^{(j)} \to 1$ or 0, the wave function is extended or localized, respectively; if $0 < \alpha_{min}^{(j)} < 1$, it is fractal.

Extending these ideas to the present case is also straightforward, with $P_l$ obtained by integrating out the photon degrees of freedom, 
leading to 
\begin{equation}\label{Eq:alpha_j}
    \alpha_{l}^{(j)}= -\frac{\log(\sum_{n}|\phi_{l,n}^{j}|^{2})}{\log( F_{m})}~.
\end{equation}
Similarly to the IPR case, it is useful to perform a multifractal analysis within a given energy window.
Then, with the same definitions as those leading to Eq.\,\eqref{eq:IPRbar}, we consider the averages
\begin{equation}
\label{Eq:alpha_avg}
    \overline{\alpha}_{min} = \frac{1}{N_{st}}\sum_{j=M_i}^{M_f}\alpha_{min}^{(j)}, 
\end{equation}
which also probe the localized, extended, or fractal character of the wavefunction, as above.

\begin{figure}[t]
    \centering
   \includegraphics[scale=0.50]{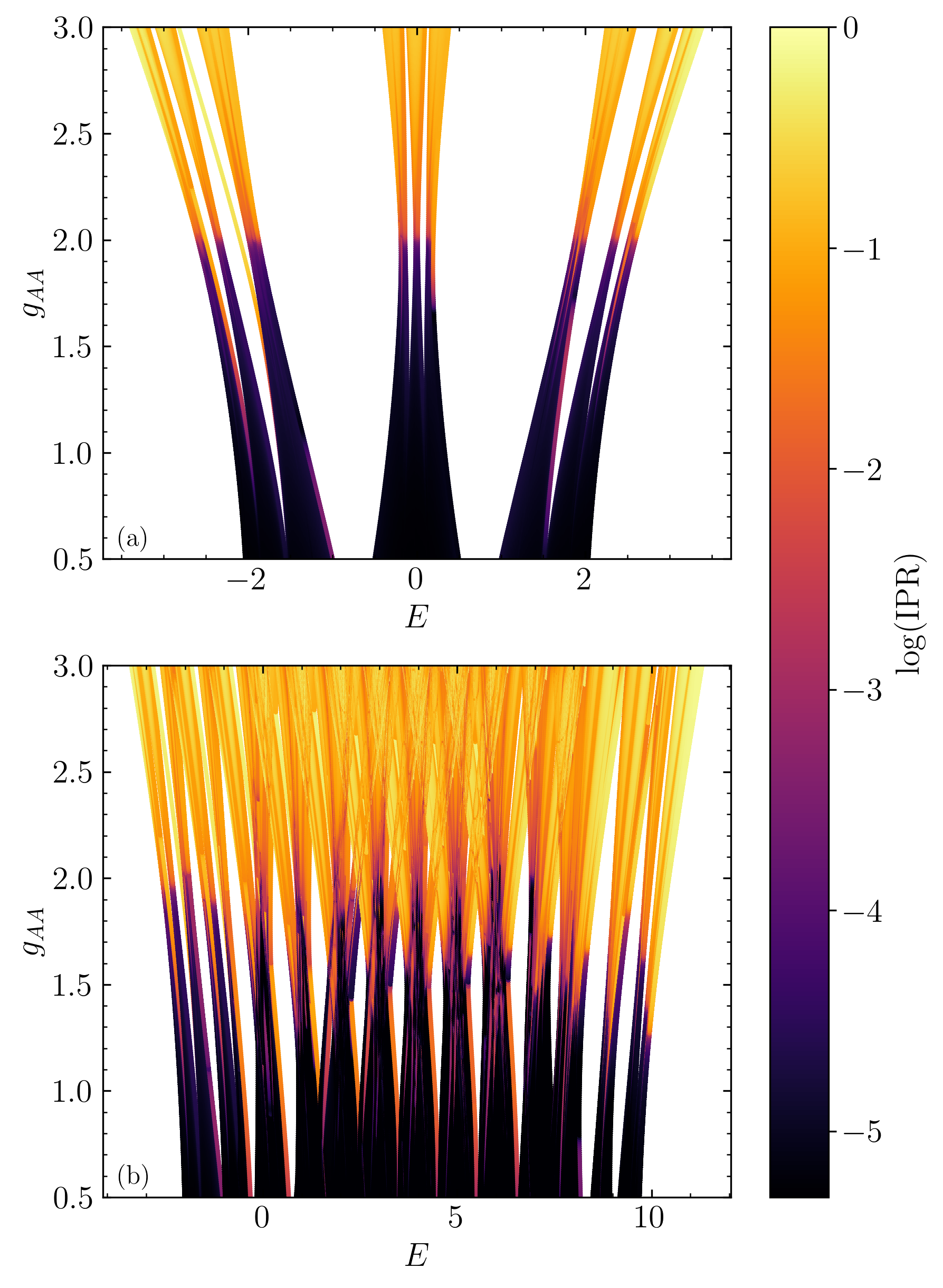}
     \caption{Numerical energy spectrum $E$ as a function of $g_{AA}$ for $L=300$ sites and fixed $\phi=0$. Panel (a) shows the standard AAH model (i.e.~no cavity effects), while panel (b) presents the case with coupling with photons. For the latter, we fixed $\omega_{0}=1$ and $\gamma = 0.15$, for $N_{ph}=8$ photons. The heatmap describes the IPR associated with each eigenstate, plotted in 10-base logarithmic scale.}
    \label{Fig:spectrum}
\end{figure}

\section{Results}
\label{Sec:results}

\begin{figure}[t]
    \centering
   \includegraphics[scale=0.40]{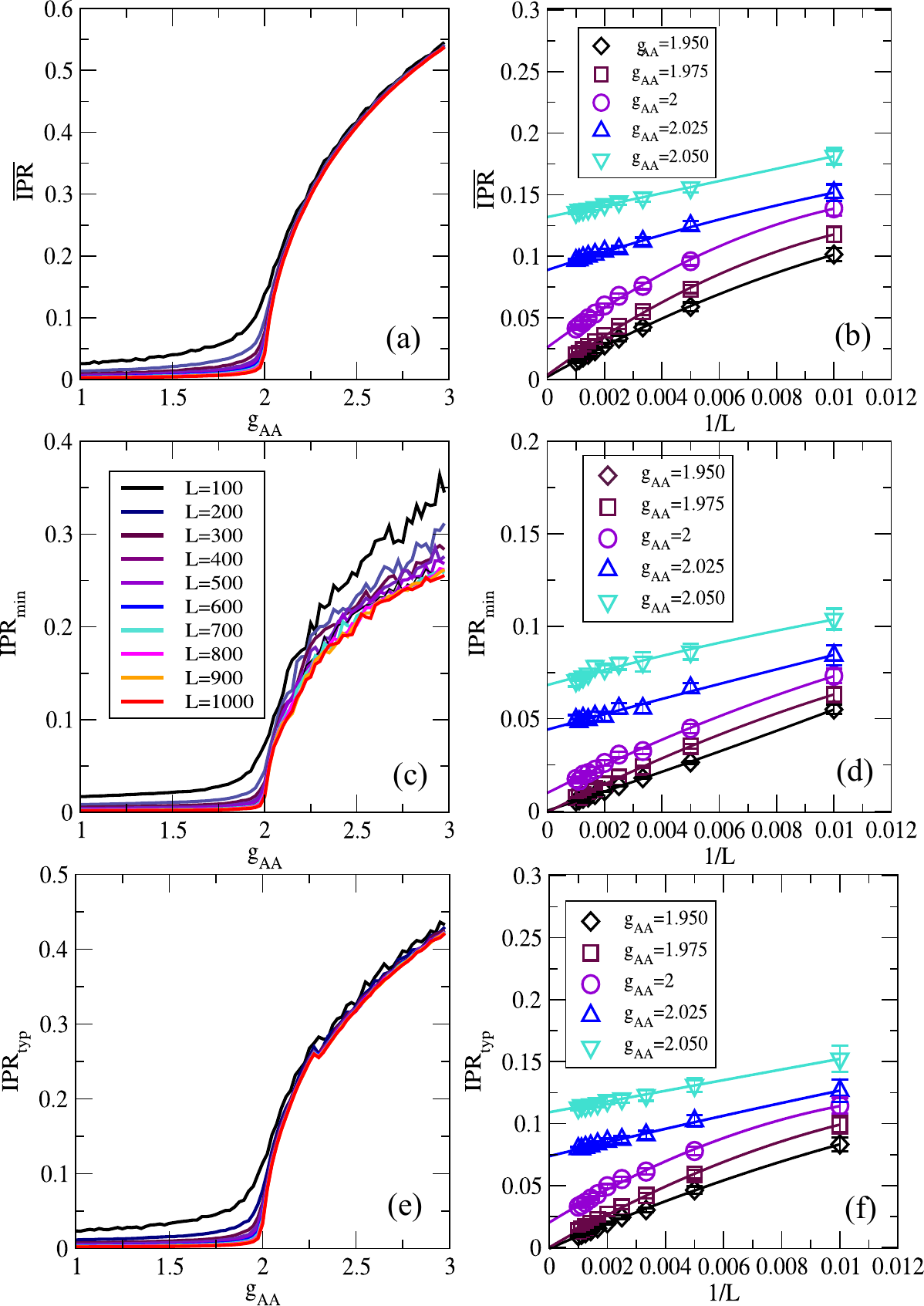}
     \caption{Inverse participation ratios for the standard AAH model. (a) $\overline{\text{IPR}}$ as a function of $g_{AA}$ for different system sizes $L$. (b) $\overline{\text{IPR}}$ as a function of $1/L$ (symbols) and extrapolated to the thermodynamic limit (solid lines). (c) and (d): same as (a) and (b), but for IPR$_{min}$. (e) and (f): same as (a) and (b), but for $\text{IPR}_{\rm typ}$.
     Here, and in all subsequent figures, when not shown, error bars are smaller than symbol sizes.}
    \label{Fig:Fig2}
\end{figure}

We begin our analysis by discussing the established properties of the AAH model without a cavity, which will serve as a baseline when discussing the effects of the coupling with photons.
Figure \ref{Fig:spectrum}\,(a) shows the spectrum of the model for different disorder strengths.
Notice that all eigenstates change their IPR near $g_{AA} = 2$,
as demanded by the self-dual property of the model. 
To probe the critical region it is more convenient to examine $\overline{\text{IPR}}$ for different system sizes, as shown in Fig.\,\ref{Fig:Fig2}\,(a), averaged over the entire spectrum.
Notice that for $g_{AA} > 2$  $\overline{\text{IPR}}$ is large and weakly dependent on the system size, whereas for $g_{AA} < 2$ it is small and decreases as $L$ increases. The critical point is obtained by employing a finite-size scaling (FSS) analysis around $g_{AA} = 2$, assuming $\overline{\text{IPR}} \propto L^{-D_f}$, as presented in Fig.\,\ref{Fig:Fig2}\,(b), from which one may notice that any $g_{AA} > 2$ leads to a finite $\overline{\text{IPR}}$.
The $\text{IPR}_{\rm typ}$ has the same behavior of the $\overline{\text{IPR}}$, as emphasized in Figs.\,\ref{Fig:Fig2}\,(e) and (f).
Similarly, one may check the behavior of $\text{IPR}_{min}$ and its FSS analysis, as shown in Figs.\,\ref{Fig:Fig2}\,(c), and (d), respectively. 
One obtains a finite response to
$\text{IPR}_{min}$ at $g_{AA} = 2$. 
These findings are consistent with our expectation that in the AAH model \textit{all} eigenstates become localized at the \textit{same} critical point.
The fractal dimension thus estimated is
$D_f = 0.53(1)$ at $g_{AA} = 2.0$, in good agreement with Ref.\,\cite{Evangelou2000}.
It is important to mention that the fractal dimension $D_f$ depends on the energy range of the eigenstates. In fact, $D_f$ exhibits a distribution around a mean value that is smaller than 1, which shows that all eigenstates exhibit energy-dependent mutifractal features at the critical point.
For instance, near the band center $E \approx 0$, the fractal dimension is $D_f \approx 0.82$, also in agreement with  Ref.\,\cite{Evangelou2000}. 

\vspace{1 cm}

\begin{figure}[t]
    \centering
    \includegraphics[scale=0.36]{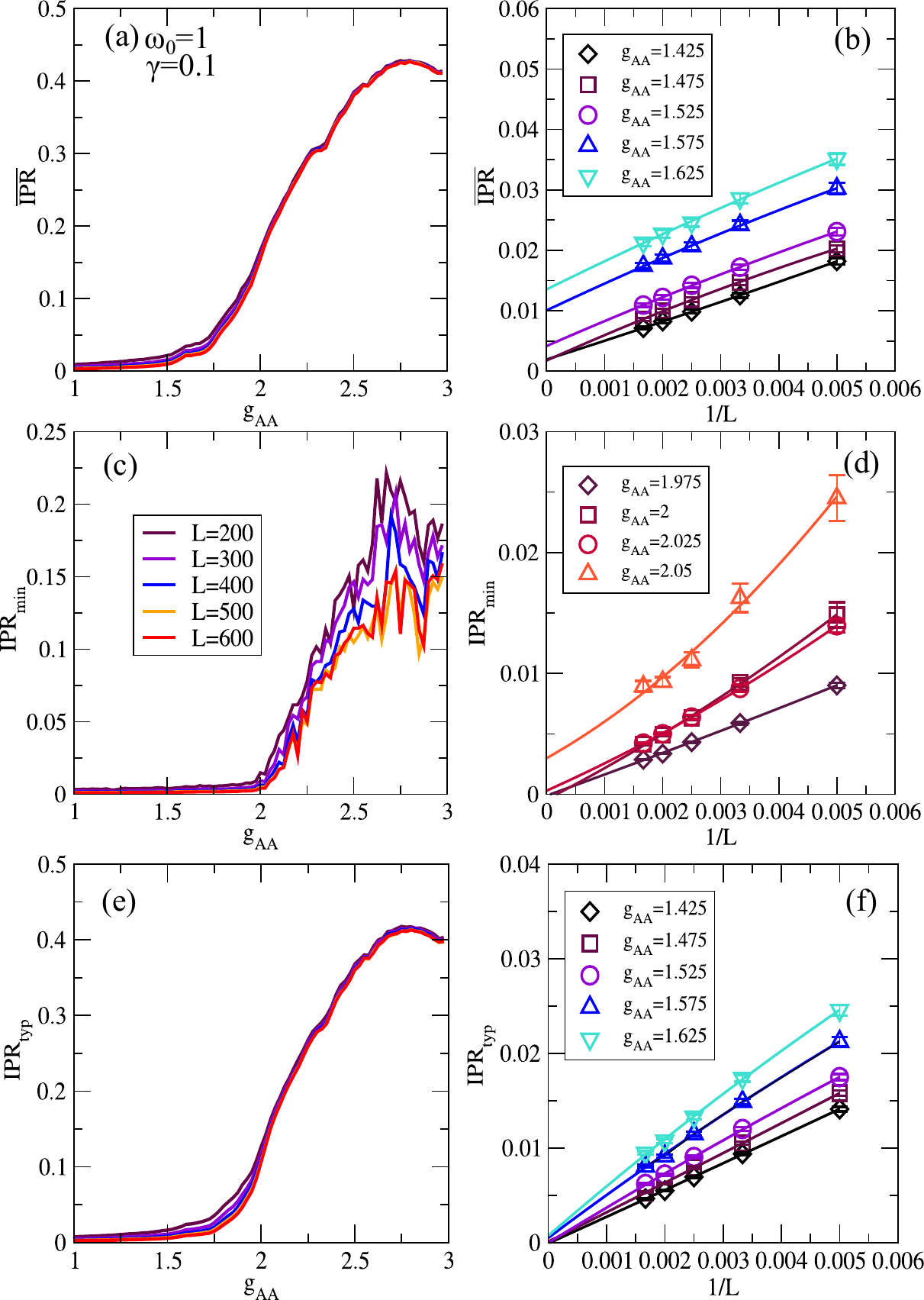} 
     \caption{Same as Fig.\,\ref{Fig:Fig2}, but for the AAH model coupled to the electromagnetic field, fixing  $ \gamma = 0.1$ and $\omega_0 = 1$. The results for the IPR are averaged over energies $E \leq 6t$.}
     \label{Fig:Fig3}
\end{figure}

We now turn to examine the cavity-coupled case. For instance, Fig.\,\ref{Fig:spectrum}\,(b) displays the energy spectrum and the IPR behavior for fixed $\omega_{0} = 1$ and $\gamma = 0.15$, from which differences with respect to the no-cavity case are apparent. 
First, the electron band is broadened due to the contribution of photons, resulting in an almost continuous spectrum. Second, although we can find extended (localized) states for $g_{AA} \ll 2$ ($g_{AA} \gg 2$), the IPR behavior is quite noisy around $g_{AA} = 2$. 
This noise is enhanced or suppressed if the coupling $\gamma$ with the electromagnetic field is increased or reduced, respectively.
At this point, a remark should be made. The broadening of the spectrum arises from the presence of different photon-number subspaces, denoted as $\hat{H}_{n,n}$ in Sec.\,\ref{subsection:model}. Each of these subspaces contains an effective copy of the AAH spectrum (with a small renormalization in the hopping integral), shifted by an energy $n \hbar \omega_0$. These subspaces are coupled by the transition terms $\hat{H}_{n,m}$ for $n \neq m$.
Therefore, when the photon energy $\hbar \omega_0$ is smaller than the bare electronic bandwidth $W \approx 4t$, energy levels from different subspaces can become degenerate. Specifically, high-energy states in $\hat{H}_{n,n}$ may become degenerate with lower-energy states in $\hat{H}_{n+1,n+1}$ or $\hat{H}_{n+2,n+2}$. As these states have distinct energy-dependent physical properties (subsumed by their fractal dimensions), their coupling with the cavity may lead to different transport properties in comparison with the uncoupled states. Indeed, as it will be discussed later, there is a clear dependence on the physical properties with the choice of $\omega_{0}$.

\begin{figure}[t]
    \centering    
    \includegraphics[scale=0.45]{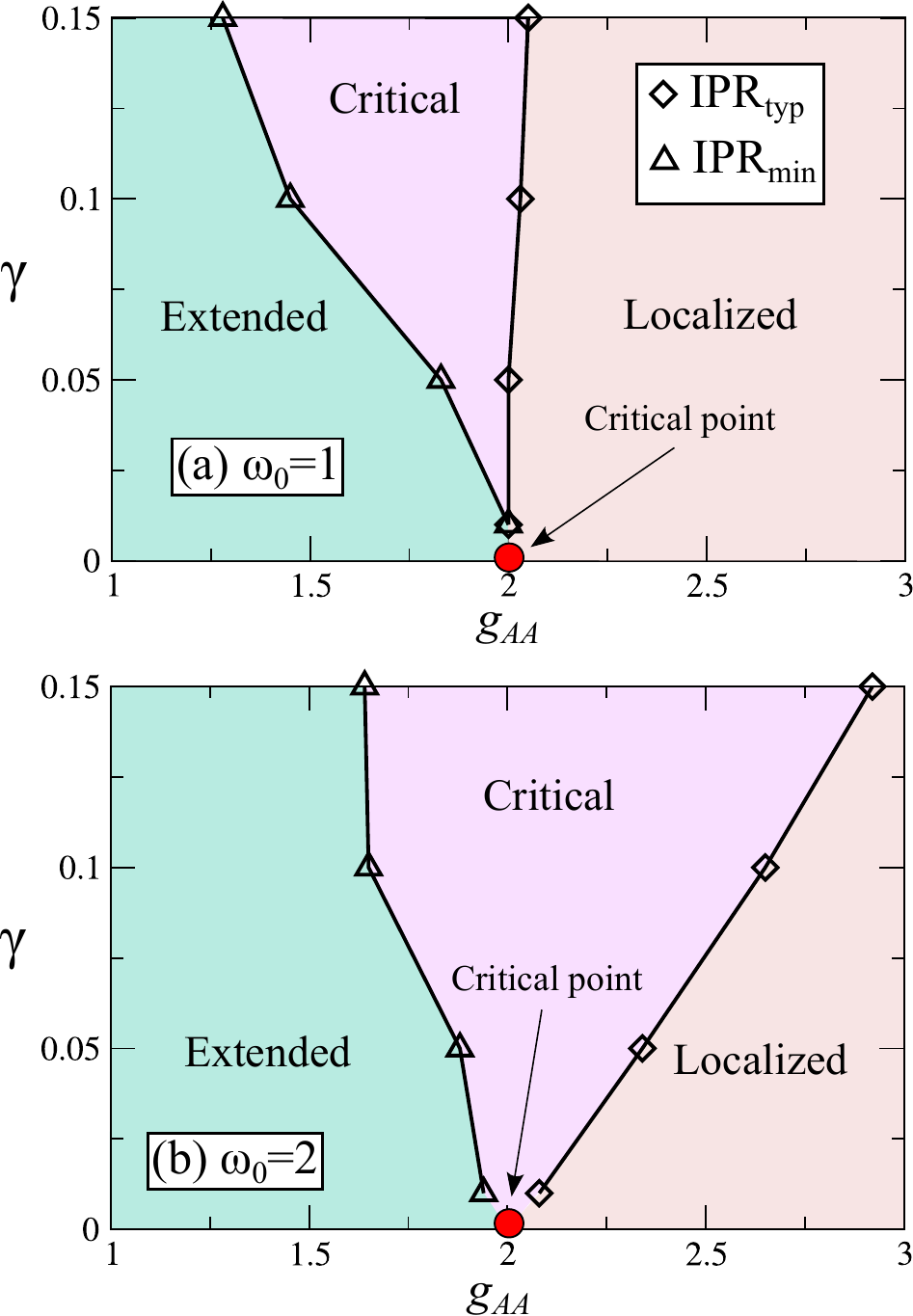}
     \caption{The phase diagram of the AAH model coupled to the electromagnetic field of the optical cavity, where we have integrated over energies $E \leq 6t$. Here, we fixed (a) $\omega_{0}=1$ and (b) $\omega_{0}=2$. The triangles denote the boundaries given by the $\text{IPR}_{\rm typ}$, while the diamonds represent those provided by the $\text{IPR}_{min}$.}
    \label{Fig:Fig7}
\end{figure}

In order to further quantify these results, we analyze  $\overline{\text{IPR}}$ (averaged over energies $E \leq 6t$) for fixed $\omega_{0}=1$ and $\gamma = 0.10$, as displayed in Fig.\,\ref{Fig:Fig3}\,(a); 
the corresponding FSS analysis is shown in Fig.\,\ref{Fig:Fig3}(b). 
Similar results for the $\text{IPR}_{\rm typ}$ are presented in Figs.\,\ref{Fig:Fig3}\,(e) and (f).
Since $\text{IPR}_{\rm typ}$ involves taking the logarithm of the IPR before averaging, it highlights the contribution of delocalized states when analyzing an inhomogeneous ensemble. This makes $\text{IPR}_{\rm typ}$ a more reliable indicator for estimating the localization transition; thus, hereafter we use it to determine the critical values.
Despite this technical nuance, both the $\overline{\text{IPR}}$ and $\text{IPR}_{\rm typ}$ yield a consistent qualitative picture: the boundary of the localized phase shifts to $g_{AA} \approx 1.6$ for $\gamma = 0.10$.
That is, the critical point shifts due to the coupling with the cavity, in stark contrast to the standard (cavity-free) case.

On the other hand, by examining $\text{IPR}_{min}$, presented in Fig.\,\ref{Fig:Fig3}\,(c), we find a small shift of it to $g_{AA} > 2$ [see, e.g., Figs.\,\ref{Fig:Fig2}\,(c)]. Figure \ref{Fig:Fig3}\,(d) displays the FSS analysis of the $\text{IPR}_{min}$, which confirms the persistence of extended states up to approximately $g_{AA} = 2.05$.
One key conclusion drawn from the previous results is the following: while $\text{IPR}_{\rm typ}$ represents the average of all IPRs, indicating when a fraction of the states ceases to be extended, $\text{IPR}_{min}$ quantifies the presence of at least one extended state. 
Thus, the former marks the boundary of the phase in which \textit{all} eigenstates are extended, while the latter marks the boundary of the phase in which \textit{all} eigenstates are localized. 
We recall that in the absence of a cavity, $\overline{\text{IPR}}$, $\text{IPR}_{\rm typ}$ and $\text{IPR}_{min}$ lead to the same critical point; see Fig.\,\ref{Fig:Fig2}.

\begin{figure}[t]
    \centering
    \includegraphics[scale=0.40]{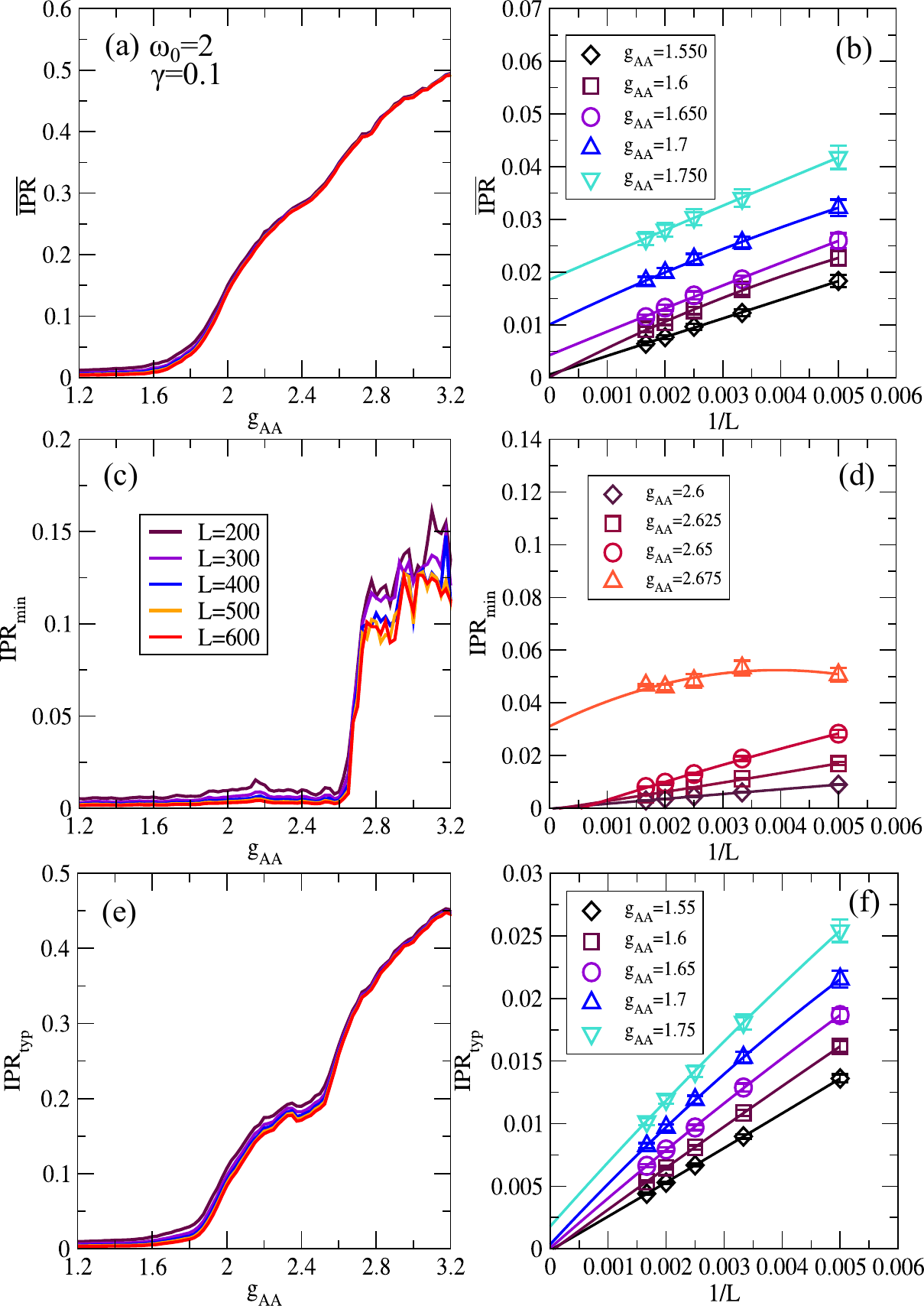}
     \caption{Same as Fig.\,\ref{Fig:Fig2}, but for the AAH model coupled to the electromagnetic field, fixing  $ \gamma = 0.1$ and $\omega_0 = 2$. The results for the IPR are averaged over energies $E \leq 6t$.}
    \label{Fig:Fig6}
\end{figure}

By repeating the same procedure outlined above for different values of $\gamma$ and fixed $\omega_{0} =1$, we obtain the phase diagram shown in Fig.\,\ref{Fig:Fig7}\,(a).
As discussed below, we anticipate that the intermediate region -- characterized by a finite $\text{IPR}_{\rm typ}$ but a vanishing $\text{IPR}_{min}$ -- contains a mixture of extended, localized, and multifractal states. Due to this coexistence of different types of wavefunction behavior, we refer to this region as \textit{critical}. However, we note that not all states within this region necessarily exhibit fractal features.
At any rate, the main effect of the electromagnetic field is to spread the critical point into a inhomogeneous intermediate region, which broadens with increasing coupling.

Let us now discuss how the photon frequency changes the shape of the phase diagram. 
Figure \ref{Fig:Fig6} exhibits the behavior of  $\overline{\text{IPR}}$, $\text{IPR}_{\rm typ}$ and $\text{IPR}_{min}$ for fixed $\omega_{0} = 2$ and $\gamma = 0.1$.
While $\overline{\text{IPR}}$ and $\text{IPR}_{\rm typ}$ are pushed to smaller values of $g_{AA}$ (similarly to the previous case), $\text{IPR}_{min}$ is pushed to larger values of $g_{AA}$, thus expanding the intermediate region.
Analyses for other values of $\gamma$ lead to the phase diagram presented in Fig.\,\ref{Fig:Fig7}\,(b), from which we see that the range of the intermediate region is broader than for $\omega_{0}=1$.
We recall that increasing the value of $\omega_0$ effectively connects high-energy states within the subspace of a given $n = n_0$ photons to low-energy states in subspaces with $n > n_0$ photons. In particular, when $\hbar \omega_0 < W$, where $W = 4t$ is the bare electronic bandwidth, energy levels from different photon-number subspaces become degenerate, and non-perturbative corrections become relevant, especially at large values of the coupling parameter $\gamma$. Since certain physical quantities -- such as the fractal dimension \cite{Evangelou2000} and diffusion coefficient \cite{Sutradhar2019} -- depend on the energy of the corresponding undressed states, this coupling between high and low-energy states can lead to significant changes, such as the broadening of the intermediate region. In other words, when degeneracies exist within the photon-number subspaces, a higher-energy photon kick can effectively overcome the system's tendency to become trapped (for some eigenstates), with the electron only localizing at a larger values of $g_{AA}$. On the other hand, if $\hbar \omega_{0} \gg W$, the absorption or emission of photons becomes unlikely, with corrections being perturbative, thus decreasing the size of the intermediate region.

\begin{figure}[t]
\includegraphics[scale=0.55]{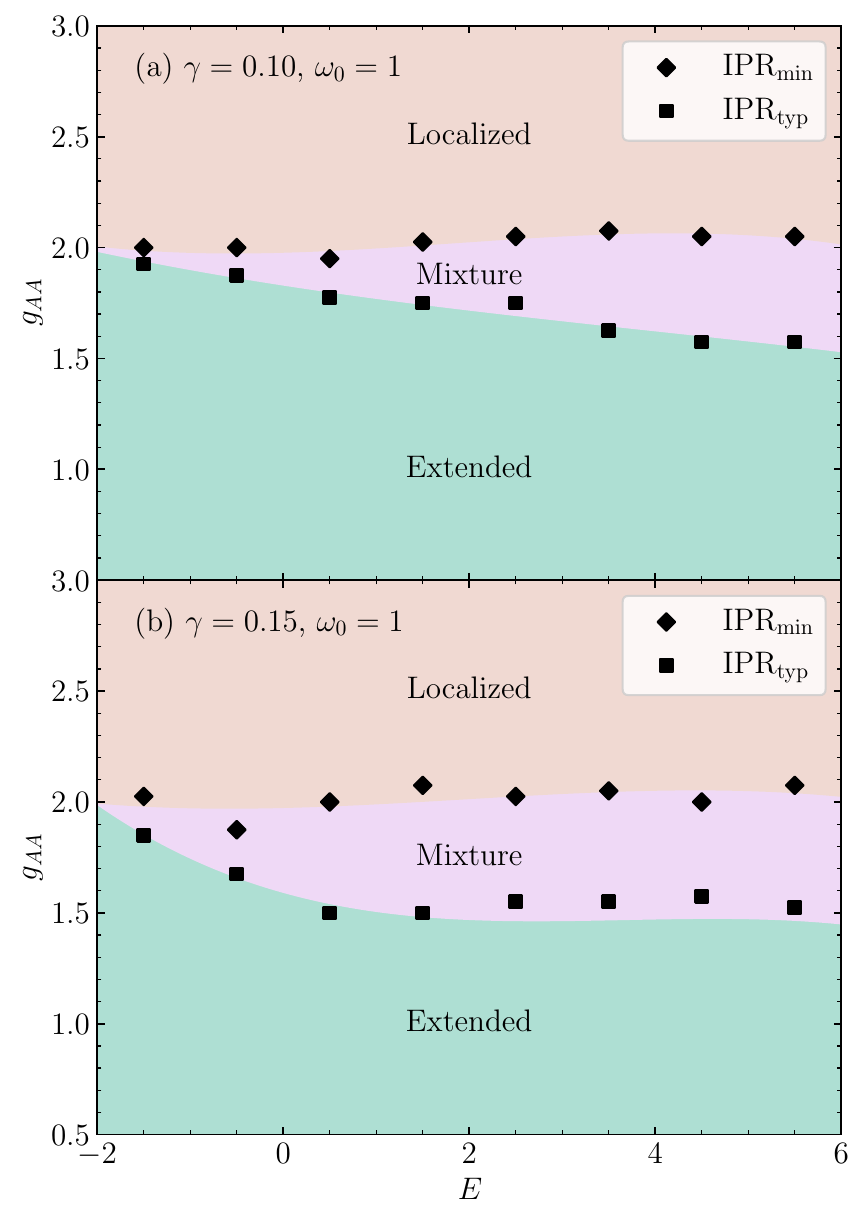}
    \caption{Energy-resolved phase diagrams of the AAH model coupled to the electromagnetic field of the optical cavity for fixed $\omega_{0}=1$, and (a) $\gamma=0.10$ and (b) $\gamma=0.15$. The square symbols denote the boundaries given by the $\text{IPR}_{\rm typ}$, while the diamonds represent those provided by the $\text{IPR}_{min}$. The hatched area are just guides to the eye.}
    \label{Fig:diagram2}
\end{figure}
 
At this point, we recall that in certain cases the kinetic energy of the electrons is well-defined, particularly in gate-induced experiments where the voltage applied to the leads can be tuned. 
Therefore, we should have at our disposal the extended-localized behavior within specific energy ranges, rather than averaging over a wider range of the spectrum.
In view of this, it is worth analyzing $\text{IPR}_{\rm typ}$ and $\text{IPR}_{min}$ at fixed energy ranges.
Here, we average both quantities over a spectrum width $|\Delta E|=~2$, centered at a given $E$.
By generically fixing $\omega_{0} = 1$, we present the energy-resolved phase diagrams in Fig.\,\ref{Fig:diagram2} for (a) $\gamma = 0.10$ and (b) $\gamma = 0.15$. 
The behavior for other values of $\gamma$ is similar, differing only by the size of the intermediate region.

There are several key points we should highlight in relation with Fig.\,\ref{Fig:diagram2}. 
First and foremost, one is still able to identify a critical region, although its width depends significantly on the position of the fixed energy interval, being wider for eigenstates with higher energies.
Thus, the boundaries in Fig.\,\ref{Fig:Fig7} should be thought of as lower and upper bounds, respectively for $g_{AA}<2$ and $g_{AA}>2$, across the energy spectrum up to $E = 6t$. 
Second, a perturbative treatment of the coupling with the field shows that transitions between states within subspaces with different number of photons are only significant if the unperturbed states are very close in energy.
Even though $\gamma=0.10$ or 0.15 lead to nonperturbative effects, the previous argument provides insights into Fig.\,\ref{Fig:diagram2}. Specifically, transitions between states in the zero-photon and one-photon subspaces are unlikely for states at the bottom of the energy spectrum, most of which remain in the zero-photon subspace, regardless of the value of $\gamma$. Consequently, for these eigenstates the effect of the coupling is less pronounced, and they are expected to follow the well-known behavior of the standard AAH model. 
This explains why the intermediate region shrinks to $g_{AA}\approx 2$ as the energy goes to the bottom of the spectrum in Fig.\,\ref{Fig:diagram2}.
Such behavior should occur for any value of coupling $\gamma$ and photon frequency $\omega_{0}$.

Now we turn to discuss the nature of the electronic states in this intermediate region. It is reasonable to suppose that the coupling with the electromagnetic field would drive the AAH critical point into a critical region. Curiously, a similar qualitative behavior occurs when dealing with spin-orbit coupling, as discussed in Ref.\,\cite{Zhou13}. Therefore, in order to examine the critical behavior of the eigenstates in the intermediate region, we now perform the multifractal analysis; see Sec.\,\ref{MFA}.
Here we deal with PBC, considering $L=F_{m}$ as a number of the Fibonacci sequence.

\begin{figure}[t]
\includegraphics[scale=0.25]{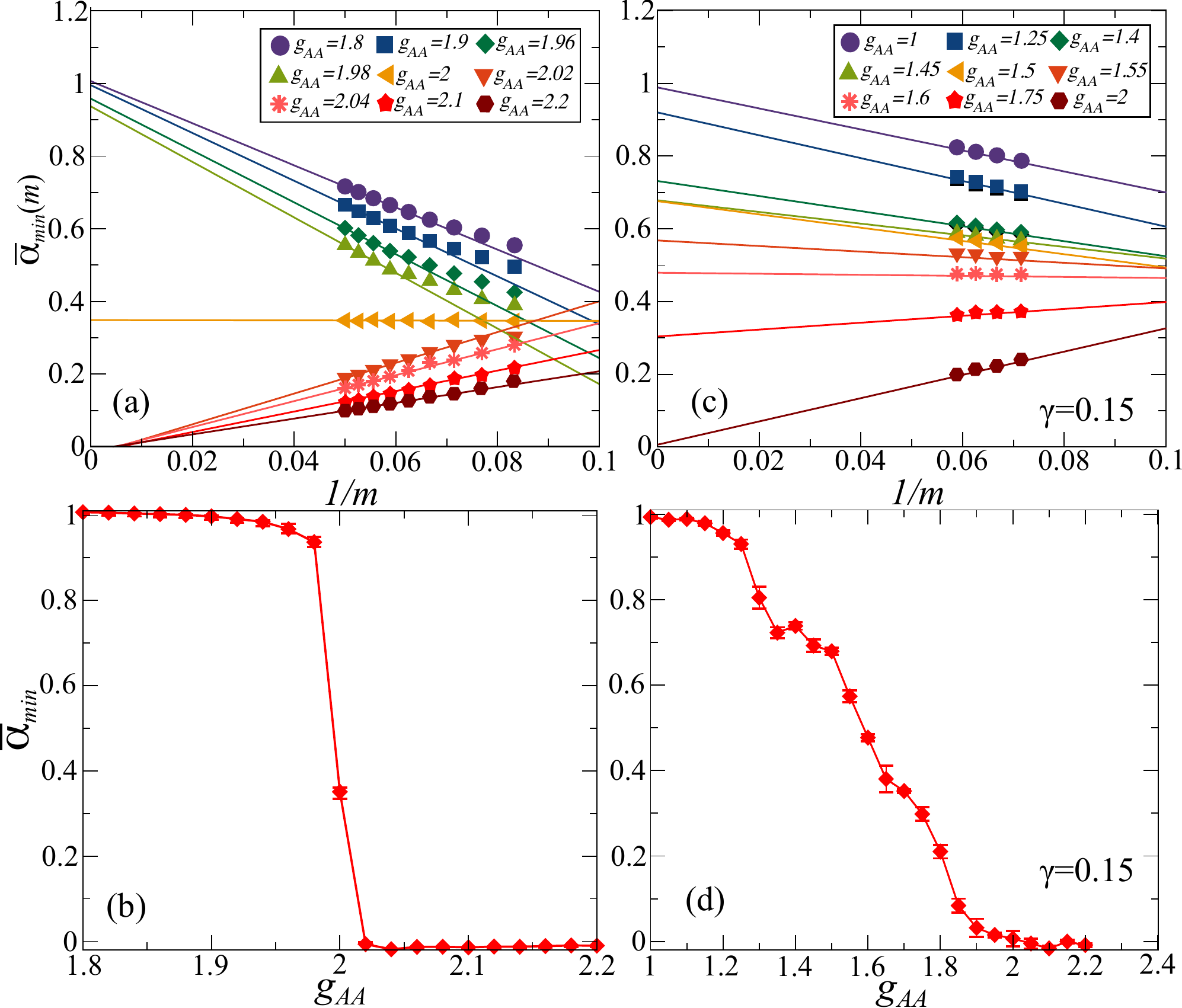}
    \caption{ Multifractal analysis for both (a)-(b) the standard AAH model and (c)-(d) the case coupled with the cavity. (a) $\overline{\alpha}_{\min}$(m) as a function of $1/m$ for different values of $g_{AA}$, close to the critical point. The solid lines denote their extrapolation to the thermodynamic limit. (b) The extrapolated results of $\overline{\alpha}_{\min}$ as a function of $g_{AA}$. Panels (c) and (d) are the same as panels (a) and (b), respectively, but for fixed $\gamma=0.15$ and $\omega_{0}=1$, while averaging over the range $\Delta E =[0,2]$.}
    \label{Fig:Fig10}
\end{figure}

To achieve this goal we start with the analysis of the usual AAH without the cavity coupling, which will serve as a reference when analyzing the effects of coupling with the electromagnetic field.
Figure \ref{Fig:Fig10}\,(a) displays the behavior of $\overline{\alpha}_{min}(m)$ as a function of $1/m$, for different system sizes, in the absence of photons \footnote{In the present case, since the phase transition of the AAH model occurs at $g_{AA}=2$ for all eigenstates, we have taken $\Delta E$ as the whole spectrum, i.e.\,we average over all eigenstates. However, we note that we have a distribution of $\alpha_{min}$ around a mean value.}.
Their extrapolated values are presented in Fig.\,\ref{Fig:Fig10}\,(b), where one may notice that, even very close to the critical point (e.g., for $g_{AA}=1.98$ or $2.02$) the extrapolations lead to $\overline{\alpha}_{min} \approx 0$ or 1. Only at $g_{AA} = 2$ one obtains an intermediate value of $\overline{\alpha}_{min}$, determining the fractal distribution of the wavefunction in the medium, and emphasizing that this is the only critical point of the standard AAH model.

Given this, we repeat the same procedure to the case with the cavity, whose results are presented in Figs.\,\ref{Fig:Fig10}\,(c) and (d), for fixed $\omega_{0}=1$ and $\gamma = 0.15$, and for different values of $g_{AA}$. As the intermediate region depends on the energy range, here we have defined $\Delta E = [0,2]$.
Notice that $\overline{\alpha}_{min} \approx 1$ when $g_{AA} \lesssim 1.2$, while $\overline{\alpha}_{min} \approx 0$ when $g_{AA} \gtrsim 2.0$, indicating extended and localized states for these regions, respectively. Indeed, the thresholds provided by  $\overline{\alpha}_{min}$ are in good agreement with the critical points identified through the $\text{IPR}_{\rm typ}$ and $\text{IPR}_{min}$ analyses, presented in Fig.\,\ref{Fig:diagram2}.
Within the intermediate region, $1.2 \lesssim g_{AA} \lesssim 2.0$, we obtain a continuous range of values for $\overline{\alpha}_{min}$, which strongly suggests the presence of critical states.
That is, differently from the standard AAH model, where such a fractal eigenstate occurs \textit{only} at the critical point $g_{AA}=2$, the coupling with the photons leads to a critical region.  

\begin{figure}[t]
\centering
\includegraphics[scale=0.55]{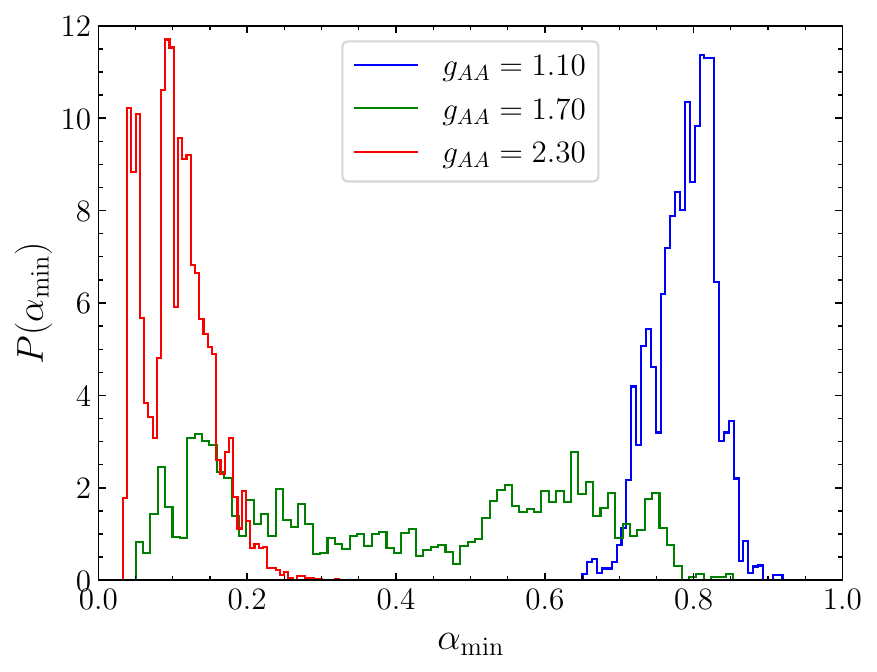}
    \caption{Probability distributions of the multifractal exponent $\alpha_{min}$ of the eigenstates within the energy range $\Delta E =[0,2]$. We fixed $\gamma=0.15$ and $\omega_{0}=1$, and examined $g_{AA}=1.1$ (large peak on the left, shown in red), $g_{AA}=1.7$ (broad distribution in the middle, shown in green), and $g_{AA}=2.3$ (large peak on the right, shown in blue). In all these cases, we examined systems with $L=987$ sites under PBC.}
    \label{Fig:P_alpha}
\end{figure}

\begin{figure*}[t]
    \centering    
    \includegraphics[scale=0.56]{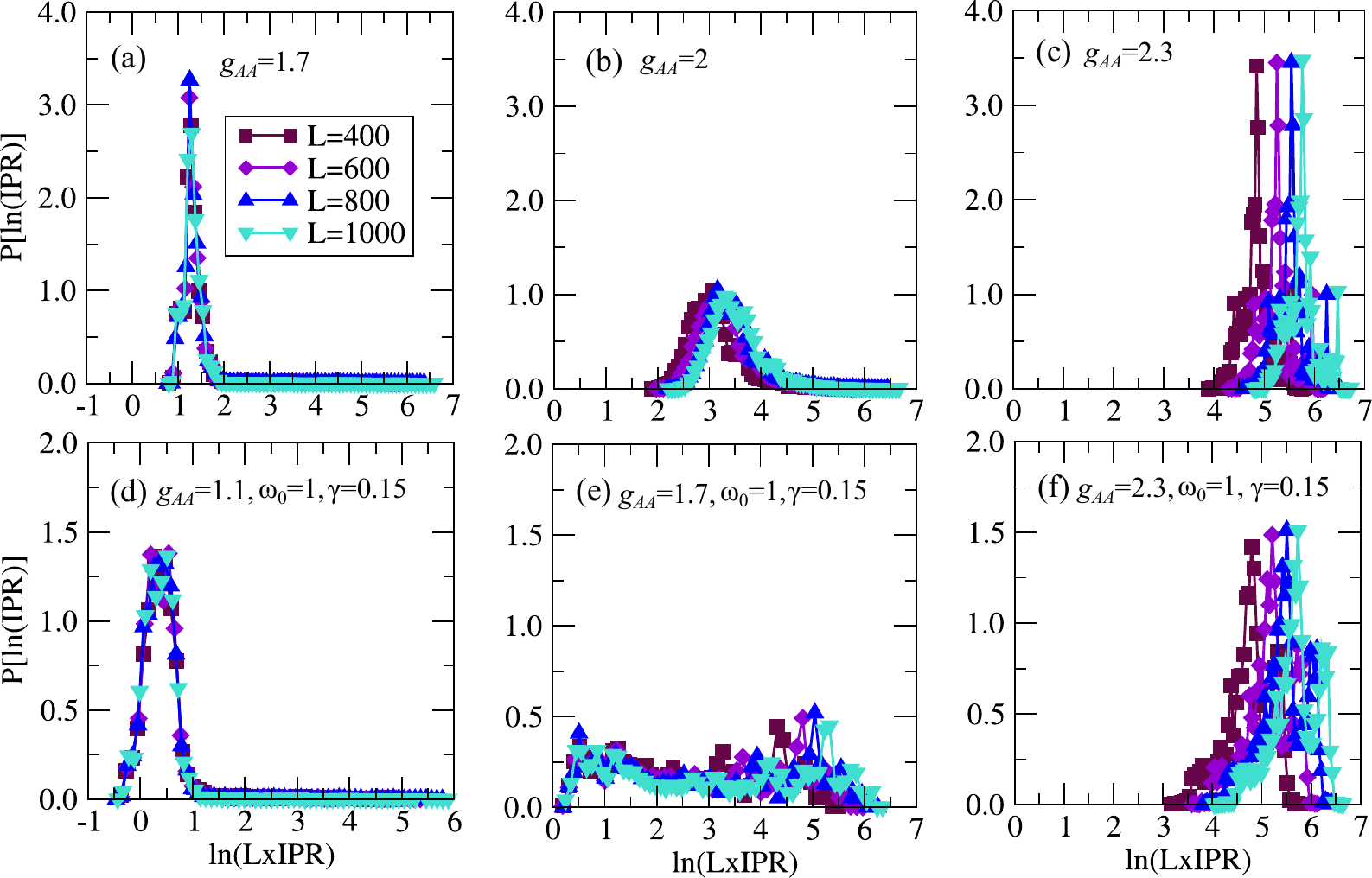}
     \caption{Probability distributions of the logarithm of the IPR as a function of $\ln \big( L \times \text{IPR} \big)$ for different system sizes. Here we examined the $P[\ln({\rm IPR})]$ for the standard AAH model (i.e.~without photons) at (a) $g_{AA}=1.7$, (b) $g_{AA}=2.0$, and (c) $g_{AA}=2.3$. Similarly, the $P[\ln({\rm IPR})]$ for the case coupled with the cavity is obtained at (d) $g_{AA}=1.1$, (e) $g_{AA}=1.7$, and (f) $g_{AA}=2.3$, for fixed $\gamma=0.15$ and $\omega_{0}=1$, while averaging over the range $\Delta E =[0,2]$.}
    \label{Fig:Fig9}
\end{figure*}

As $\overline{\alpha}_{min}$ is obtained by averaging over a given energy range, it is also appropriate to examine its probability distribution, rather than the mean value. Figure \ref{Fig:P_alpha} displays $P\big[ \alpha_{min} \big]$ for $g_{AA}=1.1$, 1.7 and 2.3, with fixed $\omega_{0}=1$, $\gamma = 0.15$, and $\Delta E = [0, 2]$. Notice that $P\big[ \alpha_{min} \big]$ exhibits a clear and sharp peak at large values of $\alpha_{min}$ for $g_{AA}=1.1$, consistent with extended states. For $g_{AA}=2.3$, the distribution also exhibits a peak, but at small values of $\alpha_{min}$, as expected for localized states. 
By contrast, fixing $g_{AA}=1.7$ within the intermediate region, we see that $P\big[ \alpha_{min} \big]$ broadens considerably. 
That is, $P\big[ \alpha_{min} \big]$ changes continuously as a function of $g_{AA}$, reshaping from a peak at large values of $\alpha_{min}$ (for large $g_{AA}$), to a peak at small values of $\alpha_{min}$ (for small $g_{AA}$), through a broad distribution in the intermediate region.
This strongly suggests a mixture of extended, localized, and \textit{critical} states within the intermediate region.

Further inspection of the previous findings can be performed by examining the probability distribution of  $\text{IPR}$, $P[\ln({\rm IPR})]$. 
In the extended regime, $L \times \text{IPR} \sim 1$, so that the maximum of the IPR distribution as a function of $\ln \big( L \times \text{IPR} \big)$ should be close to zero and independent of the system size\,\cite{Markos2006}.
In the localized regime, where $L \times \text{IPR} \propto L$, $P[\ln({\rm IPR})]$ should peak around $\ln \big( L \big)$, displaying a strong finite-size dependence.
This is observed for the standard AAH model, as shown in Figs.\,\ref{Fig:Fig9}\,(a) and (c) for the extended and localized regimes, respectively.
At the critical point of the standard AAH model ($g_{AA} = 2$), $P[\ln({\rm IPR})]$ exhibits its maximum for intermediate values of $\ln \big( L \times \text{IPR} \big)$, as presented in Fig.\,\ref{Fig:Fig9}\,(b). Notice also that $P[\ln({\rm IPR})]$ is weakly dependent on the system size at the critical point, in agreement with the dependence on the fractal dimension, $L \times \text{IPR}\propto L^{(1-D_{f})}$.

Given this, we perform the same analysis for the case with the cavity, fixing $\omega_{0}=1$, $\gamma = 0.15$, and $\Delta E = [0, 2]$. Figure \ref{Fig:Fig9}\,(d)-(f) shows the behavior of $P[\ln({\rm IPR})]$ for $g_{AA}=1.1$, $1.7$, and $2.3$, respectively. 
Similarly to the standard AAH model, panels (d) and (f) show the expected behavior for the extended and localized regimes, i.e.~with peaks at small and large values of $\ln \big( L \times \text{IPR} \big)$, respectively. 
However, in the intermediate region (e.g., for $g_{AA}=1.7$) the distribution is broader, as shown in Fig.\,\ref{Fig:Fig9}\,(e). 
Interestingly, $P[\ln({\rm IPR})]$ in panel (e) seems to exhibit contributions from size-independent peaks at small $\ln \big( L \times \text{IPR} \big)$, and also from those size-dependent at large values of $\ln \big( L \times \text{IPR} \big)$, while having a continuous distribution in between. 
This aligns with the behavior of $P\big[ \alpha_{min} \big]$, presented in Fig.\,\ref{Fig:P_alpha}, thus confirming our claim about the occurrence of a mixture of extended, localized, and critical states in the intermediate region.

\section{Conclusions}\label{conclusion}

In this paper we have investigated the effects of a single-mode electromagnetic field coupled to fermionic degrees of freedom of quasiperiodic systems. 
More specifically, we have focused on how an optical cavity affects the metal-insulator phase transition in quasiperiodic systems. 
To this end, we have considered fermions described by the one-dimensional Aubry-Andre-Harper model, which exhibits such  transition at $g_{AA}=2$ for all eigenstates; this location is exact due to the self-dual property of the model.

We have analyzed the phase transition using both the average, typical, and minimum inverse participation ratios, $\overline{\text{IPR}}$, $\text{IPR}_{\rm typ}$, and $\text{IPR}_{\text{min}}$, respectively.
While the former (average and typical) determine when a fraction of the states cease to be extended, the latter quantifies the presence of at least one extended state.
We have established that, for weak coupling with the electromagnetic cavity, the critical behavior is similar to that for the standard AAH model, with all eigenstates undergoing a phase transition at the same critical point. 
However, at intermediate or strong coupling, this single critical point broadens into an intermediate region, where  $\text{IPR}_{\text{min}}$ vanishes, but $\text{IPR}_{\rm typ}$ remains finite. 
Further, the size of this intermediate region increases with larger couplings and is strongly dependent on the photon frequency.
We have also proposed an energy-resolved phase diagram, which displays the critical points at different energy ranges. 
As the main result, we observed that the effects of coupling with the cavity are more pronounced at higher energies, with the states at the bottom of the spectrum being hardly affected.

The nature of the intermediate region was examined through a multifractal analysis, by means of the exponent $\alpha_{min}$, which provides a measure of how a given wavefunction spatially spreads through the medium. 
For the standard AAH model, it shows that the eigenstates are fractal \textit{only} at the critical point. 
A similar analysis for the field-coupled case provides evidence of fractal states within the entire intermediate region. 
We have also investigated the critical properties of the eigenstates by the probability distributions of $\alpha_{min}$ and the IPR. Both distributions are quite broad, strongly suggesting a mixture of extended, localized, and \textit{critical} states within the intermediate region. 
That is, the coupling with the electromagnetic field drives the single critical point of the standard AAH model into a critical region.

The emergence of multifractal states induced by light-matter interactions remains a subject under debate. For instance, Ref.\,\cite{Botzung2020} reports the presence of multifractal states in three-dimensional disordered arrays of quantum dots under light-induced effects. Similarly, Ref.\,\cite{Dubail2022} examines the manifestation of such states in a Bethe lattice. However, this picture changes significantly when more realistic conditions are considered. As discussed in Ref.\,\cite{Allard2022}, when dealing with a dispersive cavity and multiple cavity modes -- which are essential ingredients for accurately modeling polaritonic systems --, such multifractal states disappear. A comparable suppression is observed in interacting systems, as discussed in Ref.\,\cite{Mattiotti2024}, in which any perturbation of the integrability of the model eliminates multifractal behavior.
Therefore, the occurrence of critical states in the AAH model driven by light-matter interactions is an important step towards the understanding of metal-insulator transitions in the presence of optical cavities. 
Finally, we hope that the present results may find applications in the design of materials with specific critical transport properties, and in the study of the physics of disordered and quasicrystals structures\,\cite{Kumar2018,Xu2010, Guang2024}.

\section*{ACKNOWLEDGMENTS}
We are grateful to C.\,Lewenkopf, L.\,Martin-Moreno, T.\,Allard, G.\,Weick and C.\,Genet for valuable discussions. J.F.~thanks FAPERJ, Grant No.~SEI-260003/019642/2022;
RRdS acknowledges grants from CNPq [314611/2023-1] and FAPERJ [E-26/210.974/2024 - SEI-260003/006389/2024];
N.C.C.~acknowledges support from FAPERJ Grant No.~E-26/200.258/2023 - SEI-260003/000623/2023, CNPq Grant No.~313065/2021-7, and Serrapilheira Institute Grant No.~R-2502-52037;
T.F.M.~acknowledges support from FAPERJ Grant No.~E-26/202.518/2024 - SEI-260003/007717/2024.
F.A.P~acknowledges support from CAPES, CNPq, and FAPERJ.
Financial support from the Brazilian Agencies Conselho Nacional de Desenvolvimento Cient\'\i fico e Tecnol\'ogico (CNPq), Coordena\c c\~ao de Aperfei\c coamento de Pessoal de Ensino Superior (CAPES), and Instituto Nacional de Ci\^encia e Tecnologia de Informa\c c\~ao Qu\^antica (INCT-IQ) is also gratefully acknowledged.


\bibliography{references}

\begin{thebibliography}{64}%
\makeatletter
\providecommand \@ifxundefined [1]{%
 \@ifx{#1\undefined}
}%
\providecommand \@ifnum [1]{%
 \ifnum #1\expandafter \@firstoftwo
 \else \expandafter \@secondoftwo
 \fi
}%
\providecommand \@ifx [1]{%
 \ifx #1\expandafter \@firstoftwo
 \else \expandafter \@secondoftwo
 \fi
}%
\providecommand \natexlab [1]{#1}%
\providecommand \enquote  [1]{``#1''}%
\providecommand \bibnamefont  [1]{#1}%
\providecommand \bibfnamefont [1]{#1}%
\providecommand \citenamefont [1]{#1}%
\providecommand \href@noop [0]{\@secondoftwo}%
\providecommand \href [0]{\begingroup \@sanitize@url \@href}%
\providecommand \@href[1]{\@@startlink{#1}\@@href}%
\providecommand \@@href[1]{\endgroup#1\@@endlink}%
\providecommand \@sanitize@url [0]{\catcode `\\12\catcode `\$12\catcode
  `\&12\catcode `\#12\catcode `\^12\catcode `\_12\catcode `\%12\relax}%
\providecommand \@@startlink[1]{}%
\providecommand \@@endlink[0]{}%
\providecommand \url  [0]{\begingroup\@sanitize@url \@url }%
\providecommand \@url [1]{\endgroup\@href {#1}{\urlprefix }}%
\providecommand \urlprefix  [0]{URL }%
\providecommand \Eprint [0]{\href }%
\providecommand \doibase [0]{http://dx.doi.org/}%
\providecommand \selectlanguage [0]{\@gobble}%
\providecommand \bibinfo  [0]{\@secondoftwo}%
\providecommand \bibfield  [0]{\@secondoftwo}%
\providecommand \translation [1]{[#1]}%
\providecommand \BibitemOpen [0]{}%
\providecommand \bibitemStop [0]{}%
\providecommand \bibitemNoStop [0]{.\EOS\space}%
\providecommand \EOS [0]{\spacefactor3000\relax}%
\providecommand \BibitemShut  [1]{\csname bibitem#1\endcsname}%
\let\auto@bib@innerbib\@empty
\bibitem [{\citenamefont {Ritsch}\ \emph {et~al.}(2013)\citenamefont {Ritsch},
  \citenamefont {Domokos}, \citenamefont {Brennecke},\ and\ \citenamefont
  {Esslinger}}]{Ritsch2013}%
  \BibitemOpen
  \bibfield  {author} {\bibinfo {author} {\bibfnamefont {Helmut}\ \bibnamefont
  {Ritsch}}, \bibinfo {author} {\bibfnamefont {Peter}\ \bibnamefont {Domokos}},
  \bibinfo {author} {\bibfnamefont {Ferdinand}\ \bibnamefont {Brennecke}}, \
  and\ \bibinfo {author} {\bibfnamefont {Tilman}\ \bibnamefont {Esslinger}},\
  }\bibfield  {title} {\enquote {\bibinfo {title} {Cold atoms in
  cavity-generated dynamical optical potentials},}\ }\href {\doibase
  10.1103/RevModPhys.85.553} {\bibfield  {journal} {\bibinfo  {journal} {Rev.
  Mod. Phys.}\ }\textbf {\bibinfo {volume} {85}},\ \bibinfo {pages} {553--601}
  (\bibinfo {year} {2013})}\BibitemShut {NoStop}%
\bibitem [{\citenamefont {Schlawin}\ \emph {et~al.}(2022)\citenamefont
  {Schlawin}, \citenamefont {Kennes},\ and\ \citenamefont
  {Sentef}}]{Schlawin2022}%
  \BibitemOpen
  \bibfield  {author} {\bibinfo {author} {\bibfnamefont {F.}~\bibnamefont
  {Schlawin}}, \bibinfo {author} {\bibfnamefont {D.~M.}\ \bibnamefont
  {Kennes}}, \ and\ \bibinfo {author} {\bibfnamefont {M.~A.}\ \bibnamefont
  {Sentef}},\ }\bibfield  {title} {\enquote {\bibinfo {title} {Cavity quantum
  materials},}\ }\href {\doibase 10.1063/5.0083825} {\bibfield  {journal}
  {\bibinfo  {journal} {Appl. Phys. Rev.}\ }\textbf {\bibinfo {volume} {9}},\
  \bibinfo {pages} {011312} (\bibinfo {year} {2022})}\BibitemShut {NoStop}%
\bibitem [{\citenamefont {L{\'e}onard}\ \emph {et~al.}(2017)\citenamefont
  {L{\'e}onard}, \citenamefont {Morales}, \citenamefont {Zupancic},
  \citenamefont {Esslinger},\ and\ \citenamefont {Donner}}]{Leonard2017a}%
  \BibitemOpen
  \bibfield  {author} {\bibinfo {author} {\bibfnamefont {Julian}\ \bibnamefont
  {L{\'e}onard}}, \bibinfo {author} {\bibfnamefont {Andrea}\ \bibnamefont
  {Morales}}, \bibinfo {author} {\bibfnamefont {Philip}\ \bibnamefont
  {Zupancic}}, \bibinfo {author} {\bibfnamefont {Tilman}\ \bibnamefont
  {Esslinger}}, \ and\ \bibinfo {author} {\bibfnamefont {Tobias}\ \bibnamefont
  {Donner}},\ }\bibfield  {title} {\enquote {\bibinfo {title} {Supersolid
  formation in a quantum gas breaking a continuous translational symmetry},}\
  }\href {\doibase https://doi.org/10.1038/nature21067} {\bibfield  {journal}
  {\bibinfo  {journal} {Nature}\ }\textbf {\bibinfo {volume} {543}},\ \bibinfo
  {pages} {87--90} (\bibinfo {year} {2017})}\BibitemShut {NoStop}%
\bibitem [{\citenamefont {L\'eonard}\ \emph {et~al.}(2017)\citenamefont
  {L\'eonard}, \citenamefont {Morales}, \citenamefont {Zupancic}, \citenamefont
  {Donner},\ and\ \citenamefont {Esslinger}}]{Leonard2017b}%
  \BibitemOpen
  \bibfield  {author} {\bibinfo {author} {\bibfnamefont {Julian}\ \bibnamefont
  {L\'eonard}}, \bibinfo {author} {\bibfnamefont {Andrea}\ \bibnamefont
  {Morales}}, \bibinfo {author} {\bibfnamefont {Philip}\ \bibnamefont
  {Zupancic}}, \bibinfo {author} {\bibfnamefont {Tobias}\ \bibnamefont
  {Donner}}, \ and\ \bibinfo {author} {\bibfnamefont {Tilman}\ \bibnamefont
  {Esslinger}},\ }\bibfield  {title} {\enquote {\bibinfo {title} {Monitoring
  and manipulating {H}iggs and {G}oldstone modes in a supersolid quantum
  gas},}\ }\href {\doibase 10.1126/science.aan2608} {\bibfield  {journal}
  {\bibinfo  {journal} {Science}\ }\textbf {\bibinfo {volume} {358}},\ \bibinfo
  {pages} {1415--1418} (\bibinfo {year} {2017})},\ \Eprint
  {http://arxiv.org/abs/https://www.science.org/doi/pdf/10.1126/science.aan2608}
  {https://www.science.org/doi/pdf/10.1126/science.aan2608} \BibitemShut
  {NoStop}%
\bibitem [{\citenamefont {Landig}\ \emph {et~al.}(2016)\citenamefont {Landig},
  \citenamefont {Hruby}, \citenamefont {Dogra}, \citenamefont {Landini},
  \citenamefont {Mottl}, \citenamefont {Donner},\ and\ \citenamefont
  {Esslinger}}]{Landig2016}%
  \BibitemOpen
  \bibfield  {author} {\bibinfo {author} {\bibfnamefont {Renate}\ \bibnamefont
  {Landig}}, \bibinfo {author} {\bibfnamefont {Lorenz}\ \bibnamefont {Hruby}},
  \bibinfo {author} {\bibfnamefont {Nishant}\ \bibnamefont {Dogra}}, \bibinfo
  {author} {\bibfnamefont {Manuele}\ \bibnamefont {Landini}}, \bibinfo {author}
  {\bibfnamefont {Rafael}\ \bibnamefont {Mottl}}, \bibinfo {author}
  {\bibfnamefont {Tobias}\ \bibnamefont {Donner}}, \ and\ \bibinfo {author}
  {\bibfnamefont {Tilman}\ \bibnamefont {Esslinger}},\ }\bibfield  {title}
  {\enquote {\bibinfo {title} {Quantum phases from competing short-and
  long-range interactions in an optical lattice},}\ }\href {\doibase
  https://doi.org/10.1038/nature17409} {\bibfield  {journal} {\bibinfo
  {journal} {Nature}\ }\textbf {\bibinfo {volume} {532}},\ \bibinfo {pages}
  {476--479} (\bibinfo {year} {2016})}\BibitemShut {NoStop}%
\bibitem [{\citenamefont {Klinder}\ \emph {et~al.}(2015)\citenamefont
  {Klinder}, \citenamefont {Ke\ss{}ler}, \citenamefont {Bakhtiari},
  \citenamefont {Thorwart},\ and\ \citenamefont {Hemmerich}}]{Klinder2015}%
  \BibitemOpen
  \bibfield  {author} {\bibinfo {author} {\bibfnamefont {J.}~\bibnamefont
  {Klinder}}, \bibinfo {author} {\bibfnamefont {H.}~\bibnamefont {Ke\ss{}ler}},
  \bibinfo {author} {\bibfnamefont {M.~Reza}\ \bibnamefont {Bakhtiari}},
  \bibinfo {author} {\bibfnamefont {M.}~\bibnamefont {Thorwart}}, \ and\
  \bibinfo {author} {\bibfnamefont {A.}~\bibnamefont {Hemmerich}},\ }\bibfield
  {title} {\enquote {\bibinfo {title} {Observation of a superradiant {M}ott
  insulator in the {D}icke-{H}ubbard model},}\ }\href {\doibase
  10.1103/PhysRevLett.115.230403} {\bibfield  {journal} {\bibinfo  {journal}
  {Phys. Rev. Lett.}\ }\textbf {\bibinfo {volume} {115}},\ \bibinfo {pages}
  {230403} (\bibinfo {year} {2015})}\BibitemShut {NoStop}%
\bibitem [{\citenamefont {Vaidya}\ \emph {et~al.}(2018)\citenamefont {Vaidya},
  \citenamefont {Guo}, \citenamefont {Kroeze}, \citenamefont {Ballantine},
  \citenamefont {Koll\'ar}, \citenamefont {Keeling},\ and\ \citenamefont
  {Lev}}]{Vaidya2018}%
  \BibitemOpen
  \bibfield  {author} {\bibinfo {author} {\bibfnamefont {Varun~D.}\
  \bibnamefont {Vaidya}}, \bibinfo {author} {\bibfnamefont {Yudan}\
  \bibnamefont {Guo}}, \bibinfo {author} {\bibfnamefont {Ronen~M.}\
  \bibnamefont {Kroeze}}, \bibinfo {author} {\bibfnamefont {Kyle~E.}\
  \bibnamefont {Ballantine}}, \bibinfo {author} {\bibfnamefont {Alicia~J.}\
  \bibnamefont {Koll\'ar}}, \bibinfo {author} {\bibfnamefont {Jonathan}\
  \bibnamefont {Keeling}}, \ and\ \bibinfo {author} {\bibfnamefont
  {Benjamin~L.}\ \bibnamefont {Lev}},\ }\bibfield  {title} {\enquote {\bibinfo
  {title} {Tunable-range, photon-mediated atomic interactions in multimode
  cavity {QED}},}\ }\href {\doibase 10.1103/PhysRevX.8.011002} {\bibfield
  {journal} {\bibinfo  {journal} {Phys. Rev. X}\ }\textbf {\bibinfo {volume}
  {8}},\ \bibinfo {pages} {011002} (\bibinfo {year} {2018})}\BibitemShut
  {NoStop}%
\bibitem [{\citenamefont {Budden}\ \emph {et~al.}(2021)\citenamefont {Budden},
  \citenamefont {Gebert}, \citenamefont {Buzzi}, \citenamefont {Jotzu},
  \citenamefont {Wang}, \citenamefont {Matsuyama}, \citenamefont {Meier},
  \citenamefont {Laplace}, \citenamefont {Pontiroli}, \citenamefont
  {Ricc{\`o}}, \citenamefont {Schlawin}, \citenamefont {Jaksch},\ and\
  \citenamefont {Cavalleri}}]{Budden2021}%
  \BibitemOpen
  \bibfield  {author} {\bibinfo {author} {\bibfnamefont {M}~\bibnamefont
  {Budden}}, \bibinfo {author} {\bibfnamefont {T}~\bibnamefont {Gebert}},
  \bibinfo {author} {\bibfnamefont {M}~\bibnamefont {Buzzi}}, \bibinfo {author}
  {\bibfnamefont {G}~\bibnamefont {Jotzu}}, \bibinfo {author} {\bibfnamefont
  {E}~\bibnamefont {Wang}}, \bibinfo {author} {\bibfnamefont {T}~\bibnamefont
  {Matsuyama}}, \bibinfo {author} {\bibfnamefont {G}~\bibnamefont {Meier}},
  \bibinfo {author} {\bibfnamefont {Y}~\bibnamefont {Laplace}}, \bibinfo
  {author} {\bibfnamefont {D}~\bibnamefont {Pontiroli}}, \bibinfo {author}
  {\bibfnamefont {M}~\bibnamefont {Ricc{\`o}}}, \bibinfo {author}
  {\bibfnamefont {F.}~\bibnamefont {Schlawin}}, \bibinfo {author}
  {\bibfnamefont {D.}~\bibnamefont {Jaksch}}, \ and\ \bibinfo {author}
  {\bibfnamefont {A.}~\bibnamefont {Cavalleri}},\ }\bibfield  {title} {\enquote
  {\bibinfo {title} {Evidence for metastable photo-induced superconductivity in
  {K}3{C}60},}\ }\href {\doibase https://doi.org/10.1038/s41567-020-01148-1}
  {\bibfield  {journal} {\bibinfo  {journal} {Nature Physics}\ }\textbf
  {\bibinfo {volume} {17}},\ \bibinfo {pages} {611--618} (\bibinfo {year}
  {2021})}\BibitemShut {NoStop}%
\bibitem [{\citenamefont {Sentef}\ \emph {et~al.}(2018)\citenamefont {Sentef},
  \citenamefont {Ruggenthaler},\ and\ \citenamefont {Rubio}}]{Sentef2018}%
  \BibitemOpen
  \bibfield  {author} {\bibinfo {author} {\bibfnamefont {M.~A.}\ \bibnamefont
  {Sentef}}, \bibinfo {author} {\bibfnamefont {M.}~\bibnamefont
  {Ruggenthaler}}, \ and\ \bibinfo {author} {\bibfnamefont {A.}~\bibnamefont
  {Rubio}},\ }\bibfield  {title} {\enquote {\bibinfo {title} {Cavity
  quantum-electrodynamical polaritonically enhanced electron-phonon coupling
  and its influence on superconductivity},}\ }\href {\doibase
  10.1126/sciadv.aau6969} {\bibfield  {journal} {\bibinfo  {journal} {Science
  Advances}\ }\textbf {\bibinfo {volume} {4}},\ \bibinfo {pages} {eaau6969}
  (\bibinfo {year} {2018})},\ \Eprint
  {http://arxiv.org/abs/https://www.science.org/doi/pdf/10.1126/sciadv.aau6969}
  {https://www.science.org/doi/pdf/10.1126/sciadv.aau6969} \BibitemShut
  {NoStop}%
\bibitem [{\citenamefont {Schlawin}\ \emph {et~al.}(2019)\citenamefont
  {Schlawin}, \citenamefont {Cavalleri},\ and\ \citenamefont
  {Jaksch}}]{Schlawin2019}%
  \BibitemOpen
  \bibfield  {author} {\bibinfo {author} {\bibfnamefont {F.}~\bibnamefont
  {Schlawin}}, \bibinfo {author} {\bibfnamefont {A.}~\bibnamefont {Cavalleri}},
  \ and\ \bibinfo {author} {\bibfnamefont {D.}~\bibnamefont {Jaksch}},\
  }\bibfield  {title} {\enquote {\bibinfo {title} {Cavity-mediated
  electron-photon superconductivity},}\ }\href {\doibase
  10.1103/PhysRevLett.122.133602} {\bibfield  {journal} {\bibinfo  {journal}
  {Phys. Rev. Lett.}\ }\textbf {\bibinfo {volume} {122}},\ \bibinfo {pages}
  {133602} (\bibinfo {year} {2019})}\BibitemShut {NoStop}%
\bibitem [{\citenamefont {Curtis}\ \emph {et~al.}(2019)\citenamefont {Curtis},
  \citenamefont {Raines}, \citenamefont {Allocca}, \citenamefont {Hafezi},\
  and\ \citenamefont {Galitski}}]{Curtis2019}%
  \BibitemOpen
  \bibfield  {author} {\bibinfo {author} {\bibfnamefont {Jonathan~B.}\
  \bibnamefont {Curtis}}, \bibinfo {author} {\bibfnamefont {Zachary~M.}\
  \bibnamefont {Raines}}, \bibinfo {author} {\bibfnamefont {Andrew~A.}\
  \bibnamefont {Allocca}}, \bibinfo {author} {\bibfnamefont {Mohammad}\
  \bibnamefont {Hafezi}}, \ and\ \bibinfo {author} {\bibfnamefont {Victor~M.}\
  \bibnamefont {Galitski}},\ }\bibfield  {title} {\enquote {\bibinfo {title}
  {Cavity quantum eliashberg enhancement of superconductivity},}\ }\href
  {\doibase 10.1103/PhysRevLett.122.167002} {\bibfield  {journal} {\bibinfo
  {journal} {Phys. Rev. Lett.}\ }\textbf {\bibinfo {volume} {122}},\ \bibinfo
  {pages} {167002} (\bibinfo {year} {2019})}\BibitemShut {NoStop}%
\bibitem [{\citenamefont {Hartmann}\ \emph {et~al.}(2006)\citenamefont
  {Hartmann}, \citenamefont {Brandao},\ and\ \citenamefont
  {Plenio}}]{Hartmann2006}%
  \BibitemOpen
  \bibfield  {author} {\bibinfo {author} {\bibfnamefont {Michael~J}\
  \bibnamefont {Hartmann}}, \bibinfo {author} {\bibfnamefont {Fernando~GSL}\
  \bibnamefont {Brandao}}, \ and\ \bibinfo {author} {\bibfnamefont {Martin~B}\
  \bibnamefont {Plenio}},\ }\bibfield  {title} {\enquote {\bibinfo {title}
  {Strongly interacting polaritons in coupled arrays of cavities},}\ }\href
  {\doibase https://doi.org/10.1038/nphys462} {\bibfield  {journal} {\bibinfo
  {journal} {Nature Physics}\ }\textbf {\bibinfo {volume} {2}},\ \bibinfo
  {pages} {849--855} (\bibinfo {year} {2006})}\BibitemShut {NoStop}%
\bibitem [{\citenamefont {Tomadin}\ and\ \citenamefont
  {Fazio}(2010)}]{Tomadin2010}%
  \BibitemOpen
  \bibfield  {author} {\bibinfo {author} {\bibfnamefont {A.}~\bibnamefont
  {Tomadin}}\ and\ \bibinfo {author} {\bibfnamefont {Rosario}\ \bibnamefont
  {Fazio}},\ }\bibfield  {title} {\enquote {\bibinfo {title} {Many-body
  phenomena in {QED}-cavity arrays},}\ }\href {\doibase
  10.1364/JOSAB.27.00A130} {\bibfield  {journal} {\bibinfo  {journal} {J. Opt.
  Soc. Am. B}\ }\textbf {\bibinfo {volume} {27}},\ \bibinfo {pages}
  {A130--A136} (\bibinfo {year} {2010})}\BibitemShut {NoStop}%
\bibitem [{\citenamefont {Baum}\ \emph {et~al.}(2022)\citenamefont {Baum},
  \citenamefont {Broman}, \citenamefont {Clarke}, \citenamefont {Costa},
  \citenamefont {Mucciaccio}, \citenamefont {Yue}, \citenamefont {Zhang},
  \citenamefont {Norman}, \citenamefont {Patton}, \citenamefont {Radulaski},\
  and\ \citenamefont {Scalettar}}]{Baum2022}%
  \BibitemOpen
  \bibfield  {author} {\bibinfo {author} {\bibfnamefont {Eli}\ \bibnamefont
  {Baum}}, \bibinfo {author} {\bibfnamefont {Amelia}\ \bibnamefont {Broman}},
  \bibinfo {author} {\bibfnamefont {Trevor}\ \bibnamefont {Clarke}}, \bibinfo
  {author} {\bibfnamefont {Natanael~C.}\ \bibnamefont {Costa}}, \bibinfo
  {author} {\bibfnamefont {Jack}\ \bibnamefont {Mucciaccio}}, \bibinfo {author}
  {\bibfnamefont {Alexander}\ \bibnamefont {Yue}}, \bibinfo {author}
  {\bibfnamefont {Yuxi}\ \bibnamefont {Zhang}}, \bibinfo {author}
  {\bibfnamefont {Victoria}\ \bibnamefont {Norman}}, \bibinfo {author}
  {\bibfnamefont {Jesse}\ \bibnamefont {Patton}}, \bibinfo {author}
  {\bibfnamefont {Marina}\ \bibnamefont {Radulaski}}, \ and\ \bibinfo {author}
  {\bibfnamefont {Richard~T.}\ \bibnamefont {Scalettar}},\ }\bibfield  {title}
  {\enquote {\bibinfo {title} {Effect of emitters on quantum state transfer in
  coupled cavity arrays},}\ }\href {\doibase 10.1103/PhysRevB.105.195429}
  {\bibfield  {journal} {\bibinfo  {journal} {Phys. Rev. B}\ }\textbf {\bibinfo
  {volume} {105}},\ \bibinfo {pages} {195429} (\bibinfo {year}
  {2022})}\BibitemShut {NoStop}%
\bibitem [{\citenamefont {Saxena}\ \emph {et~al.}(2023)\citenamefont {Saxena},
  \citenamefont {Manna}, \citenamefont {Trivedi},\ and\ \citenamefont
  {Majumdar}}]{Saxena2023}%
  \BibitemOpen
  \bibfield  {author} {\bibinfo {author} {\bibfnamefont {Abhi}\ \bibnamefont
  {Saxena}}, \bibinfo {author} {\bibfnamefont {Arnab}\ \bibnamefont {Manna}},
  \bibinfo {author} {\bibfnamefont {Rahul}\ \bibnamefont {Trivedi}}, \ and\
  \bibinfo {author} {\bibfnamefont {Arka}\ \bibnamefont {Majumdar}},\
  }\bibfield  {title} {\enquote {\bibinfo {title} {Realizing tight-binding
  hamiltonians using site-controlled coupled cavity arrays},}\ }\href {\doibase
  https://doi.org/10.1038/s41467-023-41034-x} {\bibfield  {journal} {\bibinfo
  {journal} {Nature Communications}\ }\textbf {\bibinfo {volume} {14}},\
  \bibinfo {pages} {5260} (\bibinfo {year} {2023})}\BibitemShut {NoStop}%
\bibitem [{\citenamefont {Patton}\ \emph {et~al.}(2024)\citenamefont {Patton},
  \citenamefont {Norman}, \citenamefont {Mann}, \citenamefont {Puri},
  \citenamefont {Scalettar},\ and\ \citenamefont {Radulaski}}]{Patton2024}%
  \BibitemOpen
  \bibfield  {author} {\bibinfo {author} {\bibfnamefont {JT}~\bibnamefont
  {Patton}}, \bibinfo {author} {\bibfnamefont {Victoria~A}\ \bibnamefont
  {Norman}}, \bibinfo {author} {\bibfnamefont {Eliana~C}\ \bibnamefont {Mann}},
  \bibinfo {author} {\bibfnamefont {Brinda}\ \bibnamefont {Puri}}, \bibinfo
  {author} {\bibfnamefont {Richard~T}\ \bibnamefont {Scalettar}}, \ and\
  \bibinfo {author} {\bibfnamefont {Marina}\ \bibnamefont {Radulaski}},\
  }\bibfield  {title} {\enquote {\bibinfo {title} {Polariton creation in
  coupled cavity arrays with spectrally disordered emitters},}\ }\href
  {\doibase https://doi.org/10.1088/2633-4356/ad3b5d} {\bibfield  {journal}
  {\bibinfo  {journal} {Materials for Quantum Technology}\ }\textbf {\bibinfo
  {volume} {4}},\ \bibinfo {pages} {025401} (\bibinfo {year}
  {2024})}\BibitemShut {NoStop}%
\bibitem [{\citenamefont {Pisani}\ \emph {et~al.}(2023)\citenamefont {Pisani},
  \citenamefont {Gacemi},\ and\ \citenamefont {Vasanelli}}]{Pisani2023}%
  \BibitemOpen
  \bibfield  {author} {\bibinfo {author} {\bibfnamefont {F.}~\bibnamefont
  {Pisani}}, \bibinfo {author} {\bibfnamefont {D.}~\bibnamefont {Gacemi}}, \
  and\ \bibinfo {author} {\bibfnamefont {A.~et~al.}\ \bibnamefont
  {Vasanelli}},\ }\bibfield  {title} {\enquote {\bibinfo {title} {Electronic
  transport driven by collective light-matter coupled states in a quantum
  device},}\ }\href {\doibase 10.1038/s41467-023-39594-z} {\bibfield  {journal}
  {\bibinfo  {journal} {Nat Commun}\ }\textbf {\bibinfo {volume} {14}},\
  \bibinfo {pages} {3914} (\bibinfo {year} {2023})}\BibitemShut {NoStop}%
\bibitem [{\citenamefont {Appugliese}\ \emph {et~al.}(2022)\citenamefont
  {Appugliese}, \citenamefont {Enkner}, \citenamefont {Paravicini-Bagliani},
  \citenamefont {Beck}, \citenamefont {Reichl}, \citenamefont {Wegscheider},
  \citenamefont {Scalari}, \citenamefont {Ciuti},\ and\ \citenamefont
  {Faist}}]{Appugliese2022}%
  \BibitemOpen
  \bibfield  {author} {\bibinfo {author} {\bibfnamefont {F.}~\bibnamefont
  {Appugliese}}, \bibinfo {author} {\bibfnamefont {J.}~\bibnamefont {Enkner}},
  \bibinfo {author} {\bibfnamefont {G.~L.}\ \bibnamefont
  {Paravicini-Bagliani}}, \bibinfo {author} {\bibfnamefont {M.}~\bibnamefont
  {Beck}}, \bibinfo {author} {\bibfnamefont {C.}~\bibnamefont {Reichl}},
  \bibinfo {author} {\bibfnamefont {W.}~\bibnamefont {Wegscheider}}, \bibinfo
  {author} {\bibfnamefont {G.}~\bibnamefont {Scalari}}, \bibinfo {author}
  {\bibfnamefont {C.}~\bibnamefont {Ciuti}}, \ and\ \bibinfo {author}
  {\bibfnamefont {J.}~\bibnamefont {Faist}},\ }\bibfield  {title} {\enquote
  {\bibinfo {title} {Breakdown of topological protection by cavity vacuum
  fields in the integer quantum {H}all effect},}\ }\href {\doibase
  10.1126/science.abl5818} {\bibfield  {journal} {\bibinfo  {journal}
  {Science}\ }\textbf {\bibinfo {volume} {375}},\ \bibinfo {pages} {1030}
  (\bibinfo {year} {2022})}\BibitemShut {NoStop}%
\bibitem [{\citenamefont {Orgiu}\ \emph {et~al.}(2015)\citenamefont {Orgiu},
  \citenamefont {George}, \citenamefont {Hutchison}, \citenamefont {Devaux},
  \citenamefont {Dayen}, \citenamefont {Doudin}, \citenamefont {Stellacci},
  \citenamefont {Genet}, \citenamefont {Schachenmayer}, \citenamefont {Genes},
  \citenamefont {Pupillo}, \citenamefont {Samorì},\ and\ \citenamefont
  {Ebbesen}}]{Orgiu2015}%
  \BibitemOpen
  \bibfield  {author} {\bibinfo {author} {\bibfnamefont {E.}~\bibnamefont
  {Orgiu}}, \bibinfo {author} {\bibfnamefont {J.}~\bibnamefont {George}},
  \bibinfo {author} {\bibfnamefont {J.~A.}\ \bibnamefont {Hutchison}}, \bibinfo
  {author} {\bibfnamefont {E.}~\bibnamefont {Devaux}}, \bibinfo {author}
  {\bibfnamefont {J.~F.}\ \bibnamefont {Dayen}}, \bibinfo {author}
  {\bibfnamefont {B.}~\bibnamefont {Doudin}}, \bibinfo {author} {\bibfnamefont
  {F.}~\bibnamefont {Stellacci}}, \bibinfo {author} {\bibfnamefont
  {C.}~\bibnamefont {Genet}}, \bibinfo {author} {\bibfnamefont
  {J.}~\bibnamefont {Schachenmayer}}, \bibinfo {author} {\bibfnamefont
  {C.}~\bibnamefont {Genes}}, \bibinfo {author} {\bibfnamefont
  {G.}~\bibnamefont {Pupillo}}, \bibinfo {author} {\bibfnamefont
  {P.}~\bibnamefont {Samorì}}, \ and\ \bibinfo {author} {\bibfnamefont
  {T.~W.}\ \bibnamefont {Ebbesen}},\ }\bibfield  {title} {\enquote {\bibinfo
  {title} {Conductivity in organic semiconductors hybridized with the vacuum
  field},}\ }\href {\doibase 10.1038/nmat4392} {\bibfield  {journal} {\bibinfo
  {journal} {Nat. Mater.}\ }\textbf {\bibinfo {volume} {14}},\ \bibinfo {pages}
  {1123} (\bibinfo {year} {2015})}\BibitemShut {NoStop}%
\bibitem [{\citenamefont {Hagenmüller}\ \emph {et~al.}(2017)\citenamefont
  {Hagenmüller}, \citenamefont {Schachenmayer}, \citenamefont {Schütz},
  \citenamefont {Genes},\ and\ \citenamefont {Pupillo}}]{Hagenmuller2017}%
  \BibitemOpen
  \bibfield  {author} {\bibinfo {author} {\bibfnamefont {D.}~\bibnamefont
  {Hagenmüller}}, \bibinfo {author} {\bibfnamefont {J.}~\bibnamefont
  {Schachenmayer}}, \bibinfo {author} {\bibfnamefont {S.}~\bibnamefont
  {Schütz}}, \bibinfo {author} {\bibfnamefont {C.}~\bibnamefont {Genes}}, \
  and\ \bibinfo {author} {\bibfnamefont {G.}~\bibnamefont {Pupillo}},\
  }\bibfield  {title} {\enquote {\bibinfo {title} {Cavity-enhanced transport of
  charge},}\ }\href {\doibase 10.1103/PhysRevLett.119.223601} {\bibfield
  {journal} {\bibinfo  {journal} {Phys. Rev. Lett.}\ }\textbf {\bibinfo
  {volume} {119}},\ \bibinfo {pages} {223601} (\bibinfo {year}
  {2017})}\BibitemShut {NoStop}%
\bibitem [{\citenamefont {Nataf}\ \emph {et~al.}(2019)\citenamefont {Nataf},
  \citenamefont {Champel}, \citenamefont {Blatter},\ and\ \citenamefont
  {Basko}}]{Nataf2019}%
  \BibitemOpen
  \bibfield  {author} {\bibinfo {author} {\bibfnamefont {P.}~\bibnamefont
  {Nataf}}, \bibinfo {author} {\bibfnamefont {T.}~\bibnamefont {Champel}},
  \bibinfo {author} {\bibfnamefont {G.}~\bibnamefont {Blatter}}, \ and\
  \bibinfo {author} {\bibfnamefont {D.~M.}\ \bibnamefont {Basko}},\ }\bibfield
  {title} {\enquote {\bibinfo {title} {Rashba cavity {QED}: A route towards the
  superradiant quantum phase transition},}\ }\href {\doibase
  10.1103/PhysRevLett.123.207402} {\bibfield  {journal} {\bibinfo  {journal}
  {Phys. Rev. Lett.}\ }\textbf {\bibinfo {volume} {123}},\ \bibinfo {pages}
  {207402} (\bibinfo {year} {2019})}\BibitemShut {NoStop}%
\bibitem [{\citenamefont {Wang}\ \emph {et~al.}(2019)\citenamefont {Wang},
  \citenamefont {Ronca},\ and\ \citenamefont {Sentef}}]{Wang2019}%
  \BibitemOpen
  \bibfield  {author} {\bibinfo {author} {\bibfnamefont {Xiao}\ \bibnamefont
  {Wang}}, \bibinfo {author} {\bibfnamefont {Enrico}\ \bibnamefont {Ronca}}, \
  and\ \bibinfo {author} {\bibfnamefont {Michael~A.}\ \bibnamefont {Sentef}},\
  }\bibfield  {title} {\enquote {\bibinfo {title} {Cavity quantum
  electrodynamical chern insulator: Towards light-induced quantized anomalous
  {H}all effect in graphene},}\ }\href {\doibase 10.1103/PhysRevB.99.235156}
  {\bibfield  {journal} {\bibinfo  {journal} {Phys. Rev. B}\ }\textbf {\bibinfo
  {volume} {99}},\ \bibinfo {pages} {235156} (\bibinfo {year}
  {2019})}\BibitemShut {NoStop}%
\bibitem [{\citenamefont {Guerci}\ \emph {et~al.}(2020)\citenamefont {Guerci},
  \citenamefont {Simon},\ and\ \citenamefont {Mora}}]{Guerci2020}%
  \BibitemOpen
  \bibfield  {author} {\bibinfo {author} {\bibfnamefont {D.}~\bibnamefont
  {Guerci}}, \bibinfo {author} {\bibfnamefont {P.}~\bibnamefont {Simon}}, \
  and\ \bibinfo {author} {\bibfnamefont {C.}~\bibnamefont {Mora}},\ }\bibfield
  {title} {\enquote {\bibinfo {title} {Superradiant phase transition in
  electronic systems and emergent topological phases},}\ }\href {\doibase
  10.1103/PhysRevLett.125.257604} {\bibfield  {journal} {\bibinfo  {journal}
  {Phys. Rev. Lett.}\ }\textbf {\bibinfo {volume} {125}},\ \bibinfo {pages}
  {257604} (\bibinfo {year} {2020})}\BibitemShut {NoStop}%
\bibitem [{\citenamefont {Jangjan}\ and\ \citenamefont
  {Hosseini}(2020)}]{Jangjan2020}%
  \BibitemOpen
  \bibfield  {author} {\bibinfo {author} {\bibfnamefont {Milad}\ \bibnamefont
  {Jangjan}}\ and\ \bibinfo {author} {\bibfnamefont {Mir~Vahid}\ \bibnamefont
  {Hosseini}},\ }\bibfield  {title} {\enquote {\bibinfo {title} {Floquet
  engineering of topological metal states and hybridization of edge states with
  bulk states in dimerized two-leg ladders},}\ }\href@noop {} {\bibfield
  {journal} {\bibinfo  {journal} {Scientific Reports}\ }\textbf {\bibinfo
  {volume} {10}},\ \bibinfo {pages} {14256} (\bibinfo {year}
  {2020})}\BibitemShut {NoStop}%
\bibitem [{\citenamefont {Dmytruk}\ and\ \citenamefont
  {Schir{\`o}}(2022)}]{Dmytruk2022}%
  \BibitemOpen
  \bibfield  {author} {\bibinfo {author} {\bibfnamefont {Olesia}\ \bibnamefont
  {Dmytruk}}\ and\ \bibinfo {author} {\bibfnamefont {Marco}\ \bibnamefont
  {Schir{\`o}}},\ }\bibfield  {title} {\enquote {\bibinfo {title} {Controlling
  topological phases of matter with quantum light},}\ }\href@noop {} {\bibfield
   {journal} {\bibinfo  {journal} {Communications Physics}\ }\textbf {\bibinfo
  {volume} {5}},\ \bibinfo {pages} {271} (\bibinfo {year} {2022})}\BibitemShut
  {NoStop}%
\bibitem [{\citenamefont {Allard}\ and\ \citenamefont
  {Weick}(2023)}]{Allard23}%
  \BibitemOpen
  \bibfield  {author} {\bibinfo {author} {\bibfnamefont {Thomas~F.}\
  \bibnamefont {Allard}}\ and\ \bibinfo {author} {\bibfnamefont {Guillaume}\
  \bibnamefont {Weick}},\ }\bibfield  {title} {\enquote {\bibinfo {title}
  {Multiple polaritonic edge states in a {S}u-{S}chrieffer-{H}eeger chain
  strongly coupled to a multimode cavity},}\ }\href {\doibase
  10.1103/PhysRevB.108.245417} {\bibfield  {journal} {\bibinfo  {journal}
  {Phys. Rev. B}\ }\textbf {\bibinfo {volume} {108}},\ \bibinfo {pages}
  {245417} (\bibinfo {year} {2023})}\BibitemShut {NoStop}%
\bibitem [{\citenamefont {Ezawa}(2024)}]{Ezawa2024}%
  \BibitemOpen
  \bibfield  {author} {\bibinfo {author} {\bibfnamefont {Motohiko}\
  \bibnamefont {Ezawa}},\ }\bibfield  {title} {\enquote {\bibinfo {title}
  {Topological edge/corner states and polaritons in dimerized/trimerized
  superconducting qubits in a cavity},}\ }\href {\doibase
  10.1103/PhysRevB.109.205421} {\bibfield  {journal} {\bibinfo  {journal}
  {Phys. Rev. B}\ }\textbf {\bibinfo {volume} {109}},\ \bibinfo {pages}
  {205421} (\bibinfo {year} {2024})}\BibitemShut {NoStop}%
\bibitem [{\citenamefont {Liu}\ \emph {et~al.}(2023)\citenamefont {Liu},
  \citenamefont {Liu},\ and\ \citenamefont {Yao}}]{Liu2024}%
  \BibitemOpen
  \bibfield  {author} {\bibinfo {author} {\bibfnamefont {Jingyu}\ \bibnamefont
  {Liu}}, \bibinfo {author} {\bibfnamefont {Jiani}\ \bibnamefont {Liu}}, \ and\
  \bibinfo {author} {\bibfnamefont {Yao}\ \bibnamefont {Yao}},\ }\bibfield
  {title} {\enquote {\bibinfo {title} {Parity of polaritons in a molecular
  aggregate coupled to a single-mode cavity},}\ }\href {\doibase
  10.1088/1361-648X/ad1292} {\bibfield  {journal} {\bibinfo  {journal} {Journal
  of Physics: Condensed Matter}\ }\textbf {\bibinfo {volume} {36}},\ \bibinfo
  {pages} {115704} (\bibinfo {year} {2023})}\BibitemShut {NoStop}%
\bibitem [{\citenamefont {Nguyen}\ \emph {et~al.}(2024)\citenamefont {Nguyen},
  \citenamefont {Arwas},\ and\ \citenamefont {Ciuti}}]{Nguyen2024}%
  \BibitemOpen
  \bibfield  {author} {\bibinfo {author} {\bibfnamefont {Danh-Phuong}\
  \bibnamefont {Nguyen}}, \bibinfo {author} {\bibfnamefont {Geva}\ \bibnamefont
  {Arwas}}, \ and\ \bibinfo {author} {\bibfnamefont {Cristiano}\ \bibnamefont
  {Ciuti}},\ }\bibfield  {title} {\enquote {\bibinfo {title} {Electron
  conductance of a cavity-embedded topological 1d chain},}\ }\href@noop {}
  {\bibfield  {journal} {\bibinfo  {journal} {arXiv:2402.19244}\ } (\bibinfo
  {year} {2024})}\BibitemShut {NoStop}%
\bibitem [{\citenamefont {Bacciconi}\ \emph {et~al.}(2024)\citenamefont
  {Bacciconi}, \citenamefont {Andolina},\ and\ \citenamefont
  {Mora}}]{Bacciconi24}%
  \BibitemOpen
  \bibfield  {author} {\bibinfo {author} {\bibfnamefont {Zeno}\ \bibnamefont
  {Bacciconi}}, \bibinfo {author} {\bibfnamefont {Gian~Marcello}\ \bibnamefont
  {Andolina}}, \ and\ \bibinfo {author} {\bibfnamefont {Christophe}\
  \bibnamefont {Mora}},\ }\bibfield  {title} {\enquote {\bibinfo {title}
  {Topological protection of majorana polaritons in a cavity},}\ }\href
  {\doibase 10.1103/PhysRevB.109.165434} {\bibfield  {journal} {\bibinfo
  {journal} {Phys. Rev. B}\ }\textbf {\bibinfo {volume} {109}},\ \bibinfo
  {pages} {165434} (\bibinfo {year} {2024})}\BibitemShut {NoStop}%
\bibitem [{\citenamefont {G\'omez-Le\'on}\ \emph {et~al.}(2024)\citenamefont
  {G\'omez-Le\'on}, \citenamefont {Schir\'o'},\ and\ \citenamefont
  {Dmytruk}}]{Leon2024}%
  \BibitemOpen
  \bibfield  {author} {\bibinfo {author} {\bibfnamefont {\'Alvaro}\
  \bibnamefont {G\'omez-Le\'on}}, \bibinfo {author} {\bibfnamefont {Marco}\
  \bibnamefont {Schir\'o'}}, \ and\ \bibinfo {author} {\bibfnamefont {Olesia}\
  \bibnamefont {Dmytruk}},\ }\bibfield  {title} {\enquote {\bibinfo {title}
  {High-quality poor man's {M}ajorana bound states from cavity embedding},}\
  }\href@noop {} {\bibfield  {journal} {\bibinfo  {journal} {arXiv:2407.12088}\
  } (\bibinfo {year} {2024})}\BibitemShut {NoStop}%
\bibitem [{\citenamefont {Passetti}\ \emph {et~al.}(2023)\citenamefont
  {Passetti}, \citenamefont {Eckhardt}, \citenamefont {Sentef},\ and\
  \citenamefont {Kennes}}]{Passetti23}%
  \BibitemOpen
  \bibfield  {author} {\bibinfo {author} {\bibfnamefont {Giacomo}\ \bibnamefont
  {Passetti}}, \bibinfo {author} {\bibfnamefont {Christian~J.}\ \bibnamefont
  {Eckhardt}}, \bibinfo {author} {\bibfnamefont {Michael~A.}\ \bibnamefont
  {Sentef}}, \ and\ \bibinfo {author} {\bibfnamefont {Dante~M.}\ \bibnamefont
  {Kennes}},\ }\bibfield  {title} {\enquote {\bibinfo {title} {Cavity
  light-matter entanglement through quantum fluctuations},}\ }\href {\doibase
  10.1103/PhysRevLett.131.023601} {\bibfield  {journal} {\bibinfo  {journal}
  {Phys. Rev. Lett.}\ }\textbf {\bibinfo {volume} {131}},\ \bibinfo {pages}
  {023601} (\bibinfo {year} {2023})}\BibitemShut {NoStop}%
\bibitem [{\citenamefont {Ashida}\ \emph {et~al.}(2020)\citenamefont {Ashida},
  \citenamefont {Imamoglu}, \citenamefont {Faist}, \citenamefont {Jaksch},
  \citenamefont {Cavalleri},\ and\ \citenamefont {Demler}}]{Ashida2020}%
  \BibitemOpen
  \bibfield  {author} {\bibinfo {author} {\bibfnamefont {Yuto}\ \bibnamefont
  {Ashida}}, \bibinfo {author} {\bibfnamefont {Atac}\ \bibnamefont {Imamoglu}},
  \bibinfo {author} {\bibfnamefont {Jerome}\ \bibnamefont {Faist}}, \bibinfo
  {author} {\bibfnamefont {Dieter}\ \bibnamefont {Jaksch}}, \bibinfo {author}
  {\bibfnamefont {Andrea}\ \bibnamefont {Cavalleri}}, \ and\ \bibinfo {author}
  {\bibfnamefont {Eugene}\ \bibnamefont {Demler}},\ }\bibfield  {title}
  {\enquote {\bibinfo {title} {Quantum electrodynamic control of matter:
  Cavity-enhanced ferroelectric phase transition},}\ }\href {\doibase
  10.1103/PhysRevX.10.041027} {\bibfield  {journal} {\bibinfo  {journal} {Phys.
  Rev. X}\ }\textbf {\bibinfo {volume} {10}},\ \bibinfo {pages} {041027}
  (\bibinfo {year} {2020})}\BibitemShut {NoStop}%
\bibitem [{\citenamefont {Masuki}\ and\ \citenamefont
  {Ashida}(2024)}]{Masuki24}%
  \BibitemOpen
  \bibfield  {author} {\bibinfo {author} {\bibfnamefont {Kanta}\ \bibnamefont
  {Masuki}}\ and\ \bibinfo {author} {\bibfnamefont {Yuto}\ \bibnamefont
  {Ashida}},\ }\bibfield  {title} {\enquote {\bibinfo {title} {Cavity moir\'e
  materials: Controlling magnetic frustration with quantum light-matter
  interaction},}\ }\href {\doibase 10.1103/PhysRevB.109.195173} {\bibfield
  {journal} {\bibinfo  {journal} {Phys. Rev. B}\ }\textbf {\bibinfo {volume}
  {109}},\ \bibinfo {pages} {195173} (\bibinfo {year} {2024})}\BibitemShut
  {NoStop}%
\bibitem [{\citenamefont {Li}\ and\ \citenamefont {Eckstein}(2020)}]{Li2020}%
  \BibitemOpen
  \bibfield  {author} {\bibinfo {author} {\bibfnamefont {J.}~\bibnamefont
  {Li}}\ and\ \bibinfo {author} {\bibfnamefont {M.}~\bibnamefont {Eckstein}},\
  }\bibfield  {title} {\enquote {\bibinfo {title} {Manipulating intertwined
  orders in solids with quantum light},}\ }\href {\doibase
  10.1103/PhysRevLett.125.217402} {\bibfield  {journal} {\bibinfo  {journal}
  {Phys. Rev. Lett.}\ }\textbf {\bibinfo {volume} {125}},\ \bibinfo {pages}
  {217402} (\bibinfo {year} {2020})}\BibitemShut {NoStop}%
\bibitem [{\citenamefont {Mochida}\ and\ \citenamefont
  {Ashida}(2024)}]{Mochida2024}%
  \BibitemOpen
  \bibfield  {author} {\bibinfo {author} {\bibfnamefont {Jun}\ \bibnamefont
  {Mochida}}\ and\ \bibinfo {author} {\bibfnamefont {Yuto}\ \bibnamefont
  {Ashida}},\ }\bibfield  {title} {\enquote {\bibinfo {title} {Cavity-enhanced
  {K}ondo effect},}\ }\href {\doibase 10.1103/PhysRevB.110.035158} {\bibfield
  {journal} {\bibinfo  {journal} {Phys. Rev. B}\ }\textbf {\bibinfo {volume}
  {110}},\ \bibinfo {pages} {035158} (\bibinfo {year} {2024})}\BibitemShut
  {NoStop}%
\bibitem [{\citenamefont {Arwas}\ and\ \citenamefont
  {Ciuti}(2023)}]{Arwas2023}%
  \BibitemOpen
  \bibfield  {author} {\bibinfo {author} {\bibfnamefont {Geva}\ \bibnamefont
  {Arwas}}\ and\ \bibinfo {author} {\bibfnamefont {Cristiano}\ \bibnamefont
  {Ciuti}},\ }\bibfield  {title} {\enquote {\bibinfo {title} {Quantum electron
  transport controlled by cavity vacuum fields},}\ }\href {\doibase
  10.1103/PhysRevB.107.045425} {\bibfield  {journal} {\bibinfo  {journal}
  {Phys. Rev. B}\ }\textbf {\bibinfo {volume} {107}},\ \bibinfo {pages}
  {045425} (\bibinfo {year} {2023})}\BibitemShut {NoStop}%
\bibitem [{\citenamefont {Svintsov}\ \emph {et~al.}(2024)\citenamefont
  {Svintsov}, \citenamefont {Alymov}, \citenamefont {Devizorova},\ and\
  \citenamefont {Martin-Moreno}}]{Moreno2022}%
  \BibitemOpen
  \bibfield  {author} {\bibinfo {author} {\bibfnamefont {Dmitry}\ \bibnamefont
  {Svintsov}}, \bibinfo {author} {\bibfnamefont {Georgy}\ \bibnamefont
  {Alymov}}, \bibinfo {author} {\bibfnamefont {Zhanna}\ \bibnamefont
  {Devizorova}}, \ and\ \bibinfo {author} {\bibfnamefont {Luis}\ \bibnamefont
  {Martin-Moreno}},\ }\bibfield  {title} {\enquote {\bibinfo {title}
  {One-dimensional electron localization in semiconductors coupled to
  electromagnetic cavities},}\ }\href {\doibase 10.1103/PhysRevB.109.045432}
  {\bibfield  {journal} {\bibinfo  {journal} {Phys. Rev. B}\ }\textbf {\bibinfo
  {volume} {109}},\ \bibinfo {pages} {045432} (\bibinfo {year}
  {2024})}\BibitemShut {NoStop}%
\bibitem [{\citenamefont {Guo}\ and\ \citenamefont {Cai}(2024)}]{Guo2024}%
  \BibitemOpen
  \bibfield  {author} {\bibinfo {author} {\bibfnamefont {Zhengxin}\
  \bibnamefont {Guo}}\ and\ \bibinfo {author} {\bibfnamefont {Zi}~\bibnamefont
  {Cai}},\ }\bibfield  {title} {\enquote {\bibinfo {title} {Vacuum induced
  three-body delocalization in cavity quantum materials},}\ }\href@noop {}
  {\bibfield  {journal} {\bibinfo  {journal} {arXiv:2407.15032}\ } (\bibinfo
  {year} {2024})}\BibitemShut {NoStop}%
\bibitem [{\citenamefont {Harper}(1955)}]{Harper55}%
  \BibitemOpen
  \bibfield  {author} {\bibinfo {author} {\bibfnamefont {P~G}\ \bibnamefont
  {Harper}},\ }\bibfield  {title} {\enquote {\bibinfo {title} {Single band
  motion of conduction electrons in a uniform magnetic field},}\ }\href
  {\doibase 10.1088/0370-1298/68/10/304} {\bibfield  {journal} {\bibinfo
  {journal} {Proceedings of the Physical Society. Section A}\ }\textbf
  {\bibinfo {volume} {68}},\ \bibinfo {pages} {874} (\bibinfo {year}
  {1955})}\BibitemShut {NoStop}%
\bibitem [{\citenamefont {Aubry}\ and\ \citenamefont
  {Andr{\'e}}(1980)}]{Aubry80}%
  \BibitemOpen
  \bibfield  {author} {\bibinfo {author} {\bibfnamefont {Serge}\ \bibnamefont
  {Aubry}}\ and\ \bibinfo {author} {\bibfnamefont {Gilles}\ \bibnamefont
  {Andr{\'e}}},\ }\bibfield  {title} {\enquote {\bibinfo {title} {Analyticity
  breaking and {A}nderson localization in incommensurate lattices},}\
  }\href@noop {} {\bibfield  {journal} {\bibinfo  {journal} {Ann. Israel Phys.
  Soc}\ }\textbf {\bibinfo {volume} {3}},\ \bibinfo {pages} {18} (\bibinfo
  {year} {1980})}\BibitemShut {NoStop}%
\bibitem [{\citenamefont {Aulbach}\ \emph {et~al.}(2004)\citenamefont
  {Aulbach}, \citenamefont {Wobst}, \citenamefont {Ingold}, \citenamefont
  {Hänggi},\ and\ \citenamefont {Varga}}]{Aulbach2004}%
  \BibitemOpen
  \bibfield  {author} {\bibinfo {author} {\bibfnamefont {Christian}\
  \bibnamefont {Aulbach}}, \bibinfo {author} {\bibfnamefont {Andre}\
  \bibnamefont {Wobst}}, \bibinfo {author} {\bibfnamefont {Gert-Ludwig}\
  \bibnamefont {Ingold}}, \bibinfo {author} {\bibfnamefont {Peter}\
  \bibnamefont {Hänggi}}, \ and\ \bibinfo {author} {\bibfnamefont {Imre}\
  \bibnamefont {Varga}},\ }\bibfield  {title} {\enquote {\bibinfo {title}
  {Phase-space visualization of a metal–insulator transition},}\ }\href
  {\doibase 10.1088/1367-2630/6/1/070} {\bibfield  {journal} {\bibinfo
  {journal} {New J. Phys.}\ }\textbf {\bibinfo {volume} {6}},\ \bibinfo {pages}
  {70} (\bibinfo {year} {2004})}\BibitemShut {NoStop}%
\bibitem [{\citenamefont {Bu}\ \emph {et~al.}(2022)\citenamefont {Bu},
  \citenamefont {Zhai},\ and\ \citenamefont {Yin}}]{Bu2022}%
  \BibitemOpen
  \bibfield  {author} {\bibinfo {author} {\bibfnamefont {Xuan}\ \bibnamefont
  {Bu}}, \bibinfo {author} {\bibfnamefont {Liang-Jun}\ \bibnamefont {Zhai}}, \
  and\ \bibinfo {author} {\bibfnamefont {Shuai}\ \bibnamefont {Yin}},\
  }\bibfield  {title} {\enquote {\bibinfo {title} {Quantum criticality in the
  disordered {A}ubry-{A}ndr\'e model},}\ }\href {\doibase
  10.1103/PhysRevB.106.214208} {\bibfield  {journal} {\bibinfo  {journal}
  {Phys. Rev. B}\ }\textbf {\bibinfo {volume} {106}},\ \bibinfo {pages}
  {214208} (\bibinfo {year} {2022})}\BibitemShut {NoStop}%
\bibitem [{\citenamefont {Domínguez-Castro}\ and\ \citenamefont
  {Paredes}(2019)}]{Dominguez2019}%
  \BibitemOpen
  \bibfield  {author} {\bibinfo {author} {\bibfnamefont {G.~A.}\ \bibnamefont
  {Domínguez-Castro}}\ and\ \bibinfo {author} {\bibfnamefont {R.}~\bibnamefont
  {Paredes}},\ }\bibfield  {title} {\enquote {\bibinfo {title} {The
  {A}ubry–{A}ndr\'e model as a hobbyhorse for understanding the localization
  phenomenon},}\ }\href {\doibase 10.1088/1361-6404/ab1670} {\bibfield
  {journal} {\bibinfo  {journal} {Eur. J. Phys.}\ }\textbf {\bibinfo {volume}
  {40}},\ \bibinfo {pages} {045403} (\bibinfo {year} {2019})}\BibitemShut
  {NoStop}%
\bibitem [{\citenamefont {Biddle}\ \emph {et~al.}(2011)\citenamefont {Biddle},
  \citenamefont {Priour}, \citenamefont {Wang},\ and\ \citenamefont
  {Das~Sarma}}]{Biddle2011}%
  \BibitemOpen
  \bibfield  {author} {\bibinfo {author} {\bibfnamefont {J.}~\bibnamefont
  {Biddle}}, \bibinfo {author} {\bibfnamefont {D.~J.}\ \bibnamefont {Priour}},
  \bibinfo {author} {\bibfnamefont {B.}~\bibnamefont {Wang}}, \ and\ \bibinfo
  {author} {\bibfnamefont {S.}~\bibnamefont {Das~Sarma}},\ }\bibfield  {title}
  {\enquote {\bibinfo {title} {Localization in one-dimensional lattices with
  non-nearest-neighbor hopping: Generalized {A}nderson and {A}ubry-{A}ndr\'e
  models},}\ }\href {\doibase 10.1103/PhysRevB.83.075105} {\bibfield  {journal}
  {\bibinfo  {journal} {Phys. Rev. B}\ }\textbf {\bibinfo {volume} {83}},\
  \bibinfo {pages} {075105} (\bibinfo {year} {2011})}\BibitemShut {NoStop}%
\bibitem [{\citenamefont {Ganeshan}\ \emph {et~al.}(2015)\citenamefont
  {Ganeshan}, \citenamefont {Pixley},\ and\ \citenamefont
  {Das~Sarma}}]{Ganeshan2015}%
  \BibitemOpen
  \bibfield  {author} {\bibinfo {author} {\bibfnamefont {Sriram}\ \bibnamefont
  {Ganeshan}}, \bibinfo {author} {\bibfnamefont {J.~H.}\ \bibnamefont
  {Pixley}}, \ and\ \bibinfo {author} {\bibfnamefont {S.}~\bibnamefont
  {Das~Sarma}},\ }\bibfield  {title} {\enquote {\bibinfo {title} {Nearest
  neighbor tight binding models with an exact mobility edge in one
  dimension},}\ }\href {\doibase 10.1103/PhysRevLett.114.146601} {\bibfield
  {journal} {\bibinfo  {journal} {Phys. Rev. Lett.}\ }\textbf {\bibinfo
  {volume} {114}},\ \bibinfo {pages} {146601} (\bibinfo {year}
  {2015})}\BibitemShut {NoStop}%
\bibitem [{\citenamefont {Liu}\ \emph {et~al.}(2022)\citenamefont {Liu},
  \citenamefont {Xia}, \citenamefont {Longhi},\ and\ \citenamefont
  {Sanchez-Palencia}}]{Liu2022}%
  \BibitemOpen
  \bibfield  {author} {\bibinfo {author} {\bibfnamefont {Tong}\ \bibnamefont
  {Liu}}, \bibinfo {author} {\bibfnamefont {Xu}~\bibnamefont {Xia}}, \bibinfo
  {author} {\bibfnamefont {Stefano}\ \bibnamefont {Longhi}}, \ and\ \bibinfo
  {author} {\bibfnamefont {Laurent}\ \bibnamefont {Sanchez-Palencia}},\
  }\bibfield  {title} {\enquote {\bibinfo {title} {Anomalous mobility edges in
  one-dimensional quasiperiodic models},}\ }\href {\doibase
  10.21468/SciPostPhys.12.1.027} {\bibfield  {journal} {\bibinfo  {journal}
  {SciPost Phys.}\ }\textbf {\bibinfo {volume} {12}},\ \bibinfo {pages} {027}
  (\bibinfo {year} {2022})}\BibitemShut {NoStop}%
\bibitem [{\citenamefont {Rossignolo}\ and\ \citenamefont
  {Dell'Anna}(2019)}]{Rossignolo2019}%
  \BibitemOpen
  \bibfield  {author} {\bibinfo {author} {\bibfnamefont {M.}~\bibnamefont
  {Rossignolo}}\ and\ \bibinfo {author} {\bibfnamefont {L.}~\bibnamefont
  {Dell'Anna}},\ }\bibfield  {title} {\enquote {\bibinfo {title} {Localization
  transitions and mobility edges in coupled {A}ubry-{A}ndr\'e chains},}\ }\href
  {\doibase 10.1103/PhysRevB.99.054211} {\bibfield  {journal} {\bibinfo
  {journal} {Phys. Rev. B}\ }\textbf {\bibinfo {volume} {99}},\ \bibinfo
  {pages} {054211} (\bibinfo {year} {2019})}\BibitemShut {NoStop}%
\bibitem [{\citenamefont {Li}\ \emph {et~al.}(2020)\citenamefont {Li},
  \citenamefont {Golez}, \citenamefont {Mazza}, \citenamefont {Millis},
  \citenamefont {Georges},\ and\ \citenamefont {Eckstein}}]{Jiajun2020}%
  \BibitemOpen
  \bibfield  {author} {\bibinfo {author} {\bibfnamefont {Jiajun}\ \bibnamefont
  {Li}}, \bibinfo {author} {\bibfnamefont {Denis}\ \bibnamefont {Golez}},
  \bibinfo {author} {\bibfnamefont {Giacomo}\ \bibnamefont {Mazza}}, \bibinfo
  {author} {\bibfnamefont {Andrew~J.}\ \bibnamefont {Millis}}, \bibinfo
  {author} {\bibfnamefont {Antoine}\ \bibnamefont {Georges}}, \ and\ \bibinfo
  {author} {\bibfnamefont {Martin}\ \bibnamefont {Eckstein}},\ }\bibfield
  {title} {\enquote {\bibinfo {title} {Electromagnetic coupling in
  tight-binding models for strongly correlated light and matter},}\ }\href
  {\doibase 10.1103/PhysRevB.101.205140} {\bibfield  {journal} {\bibinfo
  {journal} {Phys. Rev. B}\ }\textbf {\bibinfo {volume} {101}},\ \bibinfo
  {pages} {205140} (\bibinfo {year} {2020})}\BibitemShut {NoStop}%
\bibitem [{\citenamefont {Peierls}(1933)}]{Peierls1933}%
  \BibitemOpen
  \bibfield  {author} {\bibinfo {author} {\bibfnamefont {Rudolph}\ \bibnamefont
  {Peierls}},\ }\bibfield  {title} {\enquote {\bibinfo {title} {Zur theorie des
  diamagnetismus von leitungselektronen},}\ }\href {\doibase
  https://doi.org/10.1007/BF01342591} {\bibfield  {journal} {\bibinfo
  {journal} {Zeitschrift f{\"u}r Physik}\ }\textbf {\bibinfo {volume} {80}},\
  \bibinfo {pages} {763--791} (\bibinfo {year} {1933})}\BibitemShut {NoStop}%
\bibitem [{\citenamefont {Marder}(2010)}]{Marder2010}%
  \BibitemOpen
  \bibfield  {author} {\bibinfo {author} {\bibfnamefont {Michael~P}\
  \bibnamefont {Marder}},\ }\href@noop {} {\emph {\bibinfo {title} {Condensed
  matter physics}}}\ (\bibinfo  {publisher} {John Wiley \& Sons},\ \bibinfo
  {year} {2010})\BibitemShut {NoStop}%
\bibitem [{\citenamefont {Kohmoto}(1983)}]{Kohmoto1983}%
  \BibitemOpen
  \bibfield  {author} {\bibinfo {author} {\bibfnamefont {M.}~\bibnamefont
  {Kohmoto}},\ }\bibfield  {title} {\enquote {\bibinfo {title} {Metal-insulator
  transition and scaling for incommensurate systems},}\ }\href {\doibase
  10.1103/PhysRevLett.51.1198} {\bibfield  {journal} {\bibinfo  {journal}
  {Phys. Rev. Lett.}\ }\textbf {\bibinfo {volume} {51}},\ \bibinfo {pages}
  {1198} (\bibinfo {year} {1983})}\BibitemShut {NoStop}%
\bibitem [{\citenamefont {Evangelou}\ and\ \citenamefont
  {Pichard}(2000)}]{Evangelou2000}%
  \BibitemOpen
  \bibfield  {author} {\bibinfo {author} {\bibfnamefont {S.~N.}\ \bibnamefont
  {Evangelou}}\ and\ \bibinfo {author} {\bibfnamefont {J.-L.}\ \bibnamefont
  {Pichard}},\ }\bibfield  {title} {\enquote {\bibinfo {title} {Critical
  quantum chaos and the one-dimensional {H}arper model},}\ }\href {\doibase
  10.1103/PhysRevLett.84.1643} {\bibfield  {journal} {\bibinfo  {journal}
  {Phys. Rev. Lett.}\ }\textbf {\bibinfo {volume} {84}},\ \bibinfo {pages}
  {1643--1646} (\bibinfo {year} {2000})}\BibitemShut {NoStop}%
\bibitem [{\citenamefont {Sutradhar}\ \emph {et~al.}(2019)\citenamefont
  {Sutradhar}, \citenamefont {Mukerjee}, \citenamefont {Pandit},\ and\
  \citenamefont {Banerjee}}]{Sutradhar2019}%
  \BibitemOpen
  \bibfield  {author} {\bibinfo {author} {\bibfnamefont {Jagannath}\
  \bibnamefont {Sutradhar}}, \bibinfo {author} {\bibfnamefont {Subroto}\
  \bibnamefont {Mukerjee}}, \bibinfo {author} {\bibfnamefont {Rahul}\
  \bibnamefont {Pandit}}, \ and\ \bibinfo {author} {\bibfnamefont {Sumilan}\
  \bibnamefont {Banerjee}},\ }\bibfield  {title} {\enquote {\bibinfo {title}
  {Transport, multifractality, and the breakdown of single-parameter scaling at
  the localization transition in quasiperiodic systems},}\ }\href {\doibase
  10.1103/PhysRevB.99.224204} {\bibfield  {journal} {\bibinfo  {journal} {Phys.
  Rev. B}\ }\textbf {\bibinfo {volume} {99}},\ \bibinfo {pages} {224204}
  (\bibinfo {year} {2019})}\BibitemShut {NoStop}%
\bibitem [{\citenamefont {Zhou}\ \emph {et~al.}(2013)\citenamefont {Zhou},
  \citenamefont {Pu},\ and\ \citenamefont {Zhang}}]{Zhou13}%
  \BibitemOpen
  \bibfield  {author} {\bibinfo {author} {\bibfnamefont {Lu}~\bibnamefont
  {Zhou}}, \bibinfo {author} {\bibfnamefont {Han}\ \bibnamefont {Pu}}, \ and\
  \bibinfo {author} {\bibfnamefont {Weiping}\ \bibnamefont {Zhang}},\
  }\bibfield  {title} {\enquote {\bibinfo {title} {Anderson localization of
  cold atomic gases with effective spin-orbit interaction in a quasiperiodic
  optical lattice},}\ }\href {\doibase 10.1103/PhysRevA.87.023625} {\bibfield
  {journal} {\bibinfo  {journal} {Phys. Rev. A}\ }\textbf {\bibinfo {volume}
  {87}},\ \bibinfo {pages} {023625} (\bibinfo {year} {2013})}\BibitemShut
  {NoStop}%
\bibitem [{Note1()}]{Note1}%
  \BibitemOpen
  \bibinfo {note} {In the present case, since the phase transition of the AAH
  model occurs at $g_{AA}=2$ for all eigenstates, we have taken $\Delta E$ as
  the whole spectrum, i.e.\protect \tmspace +\thinmuskip {.1667em}we average
  over all eigenstates. However, we note that we have a distribution of $\alpha
  _{min}$ around a mean value.}\BibitemShut {Stop}%
\bibitem [{\citenamefont {Markos}(2006)}]{Markos2006}%
  \BibitemOpen
  \bibfield  {author} {\bibinfo {author} {\bibfnamefont {Peter}\ \bibnamefont
  {Markos}},\ }\bibfield  {title} {\enquote {\bibinfo {title} {Numerical
  analysis of the {A}nderson localization},}\ }\href {\doibase
  10.2478/v10155-010-0081-0} {\bibfield  {journal} {\bibinfo  {journal} {Acta
  Phys. Slovaca}\ }\textbf {\bibinfo {volume} {56}},\ \bibinfo {pages}
  {561--685} (\bibinfo {year} {2006})}\BibitemShut {NoStop}%
\bibitem [{\citenamefont {Botzung}\ \emph {et~al.}(2020)\citenamefont
  {Botzung}, \citenamefont {Hagenm\"uller}, \citenamefont {Sch\"utz},
  \citenamefont {Dubail}, \citenamefont {Pupillo},\ and\ \citenamefont
  {Schachenmayer}}]{Botzung2020}%
  \BibitemOpen
  \bibfield  {author} {\bibinfo {author} {\bibfnamefont {T.}~\bibnamefont
  {Botzung}}, \bibinfo {author} {\bibfnamefont {D.}~\bibnamefont
  {Hagenm\"uller}}, \bibinfo {author} {\bibfnamefont {S.}~\bibnamefont
  {Sch\"utz}}, \bibinfo {author} {\bibfnamefont {J.}~\bibnamefont {Dubail}},
  \bibinfo {author} {\bibfnamefont {G.}~\bibnamefont {Pupillo}}, \ and\
  \bibinfo {author} {\bibfnamefont {J.}~\bibnamefont {Schachenmayer}},\
  }\bibfield  {title} {\enquote {\bibinfo {title} {Dark state semilocalization
  of quantum emitters in a cavity},}\ }\href {\doibase
  10.1103/PhysRevB.102.144202} {\bibfield  {journal} {\bibinfo  {journal}
  {Phys. Rev. B}\ }\textbf {\bibinfo {volume} {102}},\ \bibinfo {pages}
  {144202} (\bibinfo {year} {2020})}\BibitemShut {NoStop}%
\bibitem [{\citenamefont {Dubail}\ \emph {et~al.}(2022)\citenamefont {Dubail},
  \citenamefont {Botzung}, \citenamefont {Schachenmayer}, \citenamefont
  {Pupillo},\ and\ \citenamefont {Hagenm\"uller}}]{Dubail2022}%
  \BibitemOpen
  \bibfield  {author} {\bibinfo {author} {\bibfnamefont {J.}~\bibnamefont
  {Dubail}}, \bibinfo {author} {\bibfnamefont {T.}~\bibnamefont {Botzung}},
  \bibinfo {author} {\bibfnamefont {J.}~\bibnamefont {Schachenmayer}}, \bibinfo
  {author} {\bibfnamefont {G.}~\bibnamefont {Pupillo}}, \ and\ \bibinfo
  {author} {\bibfnamefont {D.}~\bibnamefont {Hagenm\"uller}},\ }\bibfield
  {title} {\enquote {\bibinfo {title} {Large random arrowhead matrices:
  Multifractality, semilocalization, and protected transport in disordered
  quantum spins coupled to a cavity},}\ }\href {\doibase
  10.1103/PhysRevA.105.023714} {\bibfield  {journal} {\bibinfo  {journal}
  {Phys. Rev. A}\ }\textbf {\bibinfo {volume} {105}},\ \bibinfo {pages}
  {023714} (\bibinfo {year} {2022})}\BibitemShut {NoStop}%
\bibitem [{\citenamefont {Allard}\ and\ \citenamefont
  {Weick}(2022)}]{Allard2022}%
  \BibitemOpen
  \bibfield  {author} {\bibinfo {author} {\bibfnamefont {Thomas~F.}\
  \bibnamefont {Allard}}\ and\ \bibinfo {author} {\bibfnamefont {Guillaume}\
  \bibnamefont {Weick}},\ }\bibfield  {title} {\enquote {\bibinfo {title}
  {Disorder-enhanced transport in a chain of lossy dipoles strongly coupled to
  cavity photons},}\ }\href {\doibase 10.1103/PhysRevB.106.245424} {\bibfield
  {journal} {\bibinfo  {journal} {Phys. Rev. B}\ }\textbf {\bibinfo {volume}
  {106}},\ \bibinfo {pages} {245424} (\bibinfo {year} {2022})}\BibitemShut
  {NoStop}%
\bibitem [{\citenamefont {Mattiotti}\ \emph {et~al.}(2024)\citenamefont
  {Mattiotti}, \citenamefont {Dubail}, \citenamefont {Hagenm\"uller},
  \citenamefont {Schachenmayer}, \citenamefont {Brantut},\ and\ \citenamefont
  {Pupillo}}]{Mattiotti2024}%
  \BibitemOpen
  \bibfield  {author} {\bibinfo {author} {\bibfnamefont {F.}~\bibnamefont
  {Mattiotti}}, \bibinfo {author} {\bibfnamefont {J.}~\bibnamefont {Dubail}},
  \bibinfo {author} {\bibfnamefont {D.}~\bibnamefont {Hagenm\"uller}}, \bibinfo
  {author} {\bibfnamefont {J.}~\bibnamefont {Schachenmayer}}, \bibinfo {author}
  {\bibfnamefont {J.-P.}\ \bibnamefont {Brantut}}, \ and\ \bibinfo {author}
  {\bibfnamefont {G.}~\bibnamefont {Pupillo}},\ }\bibfield  {title} {\enquote
  {\bibinfo {title} {Multifractality in the interacting disordered
  {T}avis-{C}ummings model},}\ }\href {\doibase 10.1103/PhysRevB.109.064202}
  {\bibfield  {journal} {\bibinfo  {journal} {Phys. Rev. B}\ }\textbf {\bibinfo
  {volume} {109}},\ \bibinfo {pages} {064202} (\bibinfo {year}
  {2024})}\BibitemShut {NoStop}%
\bibitem [{\citenamefont {Krishna~Kumar}\ \emph {et~al.}(2018)\citenamefont
  {Krishna~Kumar}, \citenamefont {Mishchenko}, \citenamefont {Chen},
  \citenamefont {Pezzini}, \citenamefont {Auton}, \citenamefont {Ponomarenko},
  \citenamefont {Zeitlerc}, \citenamefont {Eaves},\ and\ \citenamefont
  {Falko}}]{Kumar2018}%
  \BibitemOpen
  \bibfield  {author} {\bibinfo {author} {\bibfnamefont {R.}~\bibnamefont
  {Krishna~Kumar}}, \bibinfo {author} {\bibfnamefont {A.}~\bibnamefont
  {Mishchenko}}, \bibinfo {author} {\bibfnamefont {X.}~\bibnamefont {Chen}},
  \bibinfo {author} {\bibfnamefont {S.}~\bibnamefont {Pezzini}}, \bibinfo
  {author} {\bibfnamefont {G.~H.}\ \bibnamefont {Auton}}, \bibinfo {author}
  {\bibfnamefont {D.}~\bibnamefont {Ponomarenko}}, \bibinfo {author}
  {\bibfnamefont {U.}~\bibnamefont {Zeitlerc}}, \bibinfo {author}
  {\bibfnamefont {L.}~\bibnamefont {Eaves}}, \ and\ \bibinfo {author}
  {\bibfnamefont {V.~I.}\ \bibnamefont {Falko}},\ }\bibfield  {title} {\enquote
  {\bibinfo {title} {High-order fractal states in graphene superlattices},}\
  }\href {\doibase 10.1073/pnas.1804572115} {\bibfield  {journal} {\bibinfo
  {journal} {PNAS}\ }\textbf {\bibinfo {volume} {115}},\ \bibinfo {pages}
  {5135--5139} (\bibinfo {year} {2018})}\BibitemShut {NoStop}%
\bibitem [{\citenamefont {Xu}\ \emph {et~al.}(2010)\citenamefont {Xu},
  \citenamefont {Tian},\ and\ \citenamefont {Ji}}]{Xu2010}%
  \BibitemOpen
  \bibfield  {author} {\bibinfo {author} {\bibfnamefont {Pan}\ \bibnamefont
  {Xu}}, \bibinfo {author} {\bibfnamefont {HuiPing}\ \bibnamefont {Tian}}, \
  and\ \bibinfo {author} {\bibfnamefont {YueFeng}\ \bibnamefont {Ji}},\
  }\bibfield  {title} {\enquote {\bibinfo {title} {One-dimensional fractal
  photonic crystal and its characteristics},}\ }\href {\doibase
  10.1364/JOSAB.27.000640} {\bibfield  {journal} {\bibinfo  {journal} {J. Opt.
  Soc. Am. B}\ }\textbf {\bibinfo {volume} {27}},\ \bibinfo {pages} {640--647}
  (\bibinfo {year} {2010})}\BibitemShut {NoStop}%
\bibitem [{\citenamefont {Chen}\ \emph {et~al.}(2024)\citenamefont {Chen},
  \citenamefont {Lou}, \citenamefont {Hu},\ and\ \citenamefont
  {Lim}}]{Guang2024}%
  \BibitemOpen
  \bibfield  {author} {\bibinfo {author} {\bibfnamefont {Zhu-Guang}\
  \bibnamefont {Chen}}, \bibinfo {author} {\bibfnamefont {Cunzhong}\
  \bibnamefont {Lou}}, \bibinfo {author} {\bibfnamefont {Kaige}\ \bibnamefont
  {Hu}}, \ and\ \bibinfo {author} {\bibfnamefont {Lih-King}\ \bibnamefont
  {Lim}},\ }\bibfield  {title} {\enquote {\bibinfo {title} {Fractal surface
  states in three-dimensional topological quasicrystals},}\ }\href {\doibase
  10.48550/arXiv.2401.11497} {\  (\bibinfo {year} {2024}),\
  10.48550/arXiv.2401.11497},\ \Eprint {http://arxiv.org/abs/arXiv:2401.11497}
  {arXiv:arXiv:2401.11497} \BibitemShut {NoStop}%
\end{thebibliography}%

\end{document}